\documentclass[conference]{IEEEtran}
\IEEEoverridecommandlockouts

\usepackage{cite}
\usepackage{amsmath,amssymb,amsfonts}
\usepackage{algorithmic}
\usepackage{graphicx}
\usepackage{textcomp}
\usepackage{xcolor}
\usepackage[hyphens]{url}

\usepackage{wrapfig}
\usepackage{newtxmath}
\usepackage[dvipsnames]{xcolor}
\usepackage{libertine}
\usepackage[T1]{fontenc}
\usepackage{flushend}
\usepackage{fancyhdr}
\usepackage{makecell}
\usepackage{graphicx}
\usepackage{subcaption}
\usepackage{arydshln}
\usepackage[normalem]{ulem}
\usepackage{array}
\usepackage{comment}
\usepackage{listings}
\usepackage[bookmarks=true,
            breaklinks=true,
            letterpaper=true,
            colorlinks,
            citecolor=blue,
            linkcolor=blue,
            urlcolor=blue]{hyperref}
\usepackage{booktabs}
\usepackage{pifont}
\usepackage[normalem]{ulem}
\usepackage{multirow}
\usepackage{colortbl}
\usepackage{caption}
\usepackage{enumitem}
\usepackage{marginnote}
\usepackage{cleveref}

\usepackage{hyperref}
\usepackage{dirtree}
\usepackage{listings}
\usepackage{fancyvrb}
\usepackage{soul}
\usepackage{beramono}
\usepackage{balance}

\usepackage{inconsolata}

\definecolor{darkgreen}{rgb}{0.0, 0.5, 0.0} % R, G, B values
\definecolor{darkyellow}{rgb}{0.8, 0.6, 0.0} % R, G, B values

\newcommand{\cellblue}[1]{\cellcolor{blue!#1}}

\def\BibTeX{{\rm B\kern-.05em{\sc i\kern-.025em b}\kern-.08em
    T\kern-.1667em\lower.7ex\hbox{E}\kern-.125emX}}

\makeatletter
\let\oldautoref\autoref
\renewcommand{\autoref}[1]{%
  \@ifundefined{r@#1}%
    {\begingroup
      \setlength{\fboxsep}{1pt}% padding
      \colorbox{red}{\textcolor{white}{\textbf{??}}}%
     \endgroup}%
    {\oldautoref{#1}}%
}
\makeatother

\newcommand{\name}{\textsf{DICE}}
\newcommand{\hg}{\textsf{HG}}

\title{\name{}: \underline{D}etailed \underline{I}nter-\underline{C}hiplet \underline{E}nd-to-End PHY Modeling for Accurate Chiplet Simulation}

\author{\normalsize{ISCA 2026 Submission
    \textbf{\#\iscasubmissionnumber} -- Confidential Draft -- Do NOT Distribute!!}}
\makeatletter
\let\old@maketitle\@maketitle
\def\@maketitle{%
  \begin{center}
    {\small\itshape This paper has been accepted to the 53rd Annual International Symposium on Computer Architecture (ISCA 2026)\par}
    \vspace{0.5em}
  \end{center}
  \old@maketitle
}
\makeatother
\begin{document}

\newcolumntype{M}[1]{>{\centering\arraybackslash}m{#1}}

\newcommand{\eg}{\textit{e.g.}}
\newcommand{\ie}{\textit{i.e.}}
\newcommand{\etc}{\textit{etc}}
\newcommand{\vs}{\textit{vs.}}
\newcommand{\uver}{3.0}
\newcommand{\m}[1]{\(#1\)}

%%%%%%%%%%% ---CMDs---  %%%%%%%%%%%%%

\newcommand{\wip}[1]{\noindent\textcolor{blue}%
{>=================== \textsf{WIP} ====================<}}
\newcommand{\blue}[1]{\textcolor{blue}{#1}}
\newcommand{\orange}[1]{\textcolor{orange}{#1}}
\newcommand{\green}[1]{\textcolor{ForestGreen}{#1}}
\newcommand{\yellow}[1]{\textcolor{yellow}{#1}}
\newcommand{\Yes}[1]{\textcolor{black}{Yes}}
\newcommand{\No}[1]{\textcolor{black}{No}}
\newcommand{\Partial}[1]{\textcolor{black}{Partial}}
\newcommand{\Blah}[1]{\textcolor{red}{blah blah}}
\newcommand{\tv}{\vspace*{-0.5cm}}
\newcommand{\phy}{\textsf{PHY}}
\newcommand{\review}[1]{}
\newcommand{\red}[1]{{#1}}
\newcommand{\dc}[1]{{#1}}
\renewcommand{\sectionautorefname}{Section}
\renewcommand{\subsectionautorefname}{Section}
\renewcommand{\subsubsectionautorefname}{Section}

\crefname{figure}{Figure}{Figure}
\Crefname{figure}{Figure}{Figure}

%%%%%%%%%%% ---Rebuttal colors---  %%%%%%%%%%%%%

\colorlet{rebut}{black}
\DeclareRobustCommand{\rebut}[1]{\textcolor{rebut}{#1}}

\author{
\IEEEauthorblockN{Rashid Aligholipour}
\IEEEauthorblockA{
rashid.aligholipour@it.uu.se \\
Uppsala University \\
Sweden
}
\and
\IEEEauthorblockN{Stefanos Kaxiras}
\IEEEauthorblockA{
stefanos.kaxiras@it.uu.se \\
Uppsala University \\
Sweden
}
\and
\IEEEauthorblockN{Yuan Yao}
\IEEEauthorblockA{
yuan.yao@it.uu.se \\
Uppsala University \\
Sweden
}
}

\maketitle
\thispagestyle{plain}
\pagestyle{plain}
\pagenumbering{gobble}

%% EDIT YOUR PAPER'S CONTENTS BELOW

\begin{abstract}
Scaling monolithic multicores is increasingly constrained by power/thermal limits, yield, and rising manufacturing and testing costs. Chiplet designs address these challenges by partitioning large dies into smaller parts (typically multiple core-complex dies and an I/O die) linked via high-bandwidth physical fabrics (\textsf{PHY}). As bandwidth and wiring density scale, however, these short-reach links are pushed closer to their signal-integrity limits, increasing susceptibility to noise, crosstalk, and channel loss, motivating stronger link-level reliability mechanisms such as forward error correction (FEC). Despite this trend, state-of-the-art simulation infrastructures often approximate inter-chiplet links using oversimplified, fixed-latency models. Such abstractions overlook the inherently \emph{dynamic, runtime-dependent} behavior of the \textsf{PHY}---including channel conditions (\textit{e.g.}, signal-to-noise ratio shifts, signal crosstalk, clock jitter), iterative decoder convergence and packet retransmissions, and application dynamics (\textit{e.g.}, LLC-misses that travel across chiplet boundaries)---all of which are hard to determine offline. We show that neglecting these effects distorts inter-chiplet packet-level timing and high-level performance metrics such as IPC, leading to off-trend simulation results.

We present \textsf{DICE}, an \emph{in-simulation, runtime} \textsf{PHY} modeling in gem5 that captures the end-to-end inter-chiplet datapath, including QC-LDPC encoding/decoding, PAM4 modulation, lossy-channel transmission, LLR-based demodulation, adaptive packet re-sending, and \textsf{PHY}-level flow control between chiplets. For fidelity, we \emph{calibrate} component latencies via hardware synthesis (\textit{e.g.}, QC-LDPC decode paths and iteration budgets) and compare the integrated system against production chiplet processors. 
Compared with state-of-the-art fixed-latency inter-chiplet links such as \textsf{HeteroGarnet}, \textsf{DICE} reshapes packet-latency composition and shifts system IPC by an average of \dc{6.8\%} and up to \dc{27.6\%}, revealing variability driven by realistic \textsf{PHY} link behaviors, enabling more accurate co-evaluation of performance, reliability, and design trade-offs in modern chiplet systems.
\end{abstract}

\maketitle

\section{Introduction}

% As the number of cores in chip multiprocessors (CMPs) increases to enhance throughput, chip designers face several major challenges: 1) power consumption and the resulting thermal issues\cite{MI300_Micro}, 2) complications in both pre- and post-fabrication testing\cite{AMD_EPYC}, 3) running up against the lithographic reticle limit, and 4) low manufacturing yield\cite{AMD_MI300,AMD_EPYC}. These along with other boundaries make it increasingly impractical to build large monolithic chips that integrate a high number of cores \cite{AMD_EPYC, AMD_MI300}.

% This architecture, not just leveraged in the cutting-edge servers, but also widely employed in desktops, mobile or even in embedded domains. \autoref{fig:AMD_timeline} illustrated the recent AMD processors (EPYC, Ryzen and Instint series), which utilized chiplets in their architecture. This initiative approach commenced first in 2017 with AMD EPYC Naples with four chiplets and further more processors from diverse purpose adopted chiplets in their design.

\begin{figure}[!t]
\centering
    \includegraphics[width=0.9\linewidth]{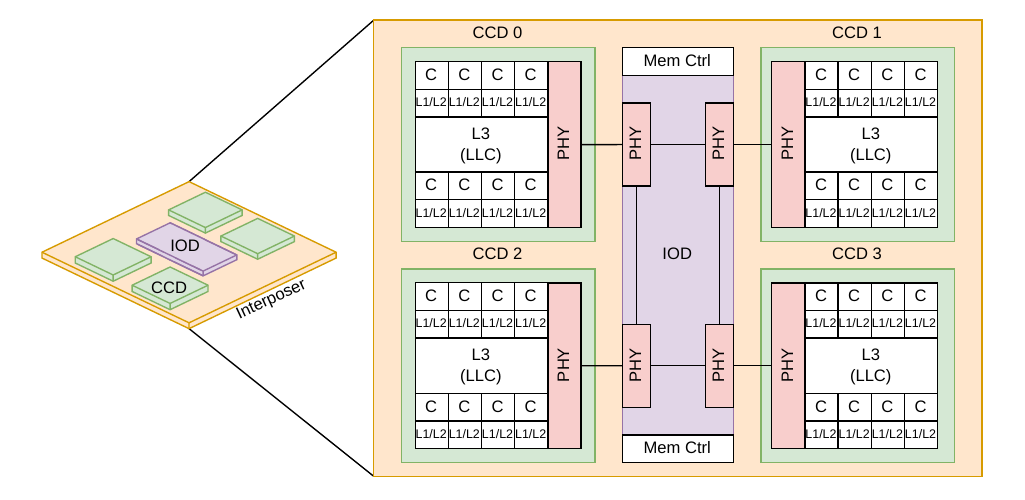}
  \caption{An example chiplet architecture, where inter-chiplet communications are done via \phy{} connects.}
  \label{fig:intro}
  \tv{}
\end{figure}

\textbf{Background.} 
As multiprocessors grow larger, keeping them monolithic has become increasingly hard. Power and thermal ceilings~\cite{MI300_Micro}, manufacturing-yield at large die sizes~\cite{Jason_photonic}, and the rising cost of testing~\cite{AMD_EPYC} all hinder further scaling. These pressures motivate \emph{chiplet} designs as a cost-efficient alternative~\cite{AMD_EPYC}, where a large monolithic die is partitioned into smaller chiplets (\autoref{fig:intro})---for example, Core Complex Dies (CCDs), which host compute cores and caches, and an I/O Die (IOD), which manages communication with off-chip DRAM and I/O devices. For communication, chiplets are interconnected via high-density physical-layer (\phy{}) links in an interposer~\cite{ECTC_France, ATTACK_CHIP_INTER}, or emerging packaging technology such as TSMC's Integrated Fan-Out on Substrate (InFO-oS), and transfer data using protocols such as AMD's Infinity Fabric~\cite{Zen4_MICRO}, Intel's Advanced Interface Bus (AIB)~\cite{Kehlet2017AIB}, and the open standard Universal Chiplet Interconnect Express (UCIe)~\cite{9893865}, \etc{}. As bandwidths rise and wiring densities increase, these short-reach links are pushed closer to their signal-integrity limits, making them more vulnerable to noise, crosstalk, and channel loss. Consequently, future inter-chiplet links are also considering the integration of forward error correction (FEC) to maintain reliable data transmission~\cite{11150380, IEEE_HIR_HPC_2021,synopsis}.

\begin{figure}[!t]
  %\begin{center}
    %\includegraphics[width=0.475\textwidth]{figs/text/yuan/overview.pdf}
    \centering
    \includegraphics[width=0.5\textwidth]{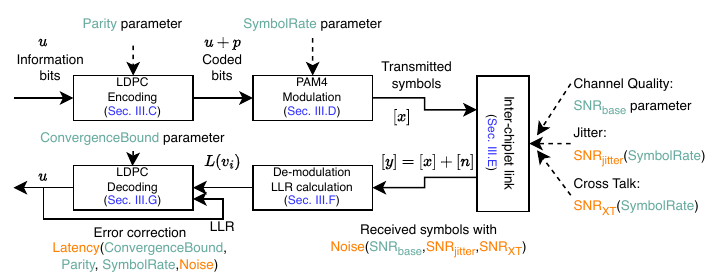}
  %\end{center}
  \caption{End-to-end inter-chiplet data communication in \name{}. Parameters in {\color{teal}green}, computed metrics in \color{orange}{orange}.}
  \label{fig:overview}
  \tv{}
\end{figure}

\textbf{The problems.}
Despite the wide adoption of chiplet-based design, most architectural simulators still ``wire up'' chiplets using interconnect models originally developed for monolithic dies, resulting in two problems. First, these models typically assume fixed link latencies, abstracting away the details of the inter-chiplet \phy{}, where parallel bits are serialized and modulated into serial signals, transmitted as waveforms, and recovered over noise-prone channels. Second, \phy{} modeling is inherently \emph{dynamic}: parity bits, symbol rate, and channel signal-to-noise ratio (SNR) jointly determine bit reliability, decoder convergence, error-correction latency, \etc{} (as illustrated in~\autoref{fig:overview}). Tuning any one parameter perturbs the others, forming a tightly coupled loop that fixed-delay abstractions cannot faithfully capture or calibrate easily. Consequently, fixed-latency links often yielding coarse---and sometimes off-trend---simulation conclusions.

\textbf{\name{}} %%% DO NOT SAY MAIN IDEA
addresses these limitations by modeling the complete inter-chiplet \phy{} pipeline. Guided by open-standard specifications such as the IEEE Heterogeneous Integration Roadmap~\cite{IEEE_HIR_HPC_2021}, \name{} incorporates LDPC encoding (\autoref{sec:fec:encode}), PAM4 modulation (\autoref{sec:ber}), lossy-channel transmission (\autoref{sec:awgn}), LLR-based demodulation (\autoref{sec:llr}), and LDPC decoding (\autoref{sec:fec:decode}). By simulating these interactions at runtime (\autoref{sec:router-micro}), \name{} captures the physical-layer effects that significantly influence architecture-level simulations.

\begin{figure}[!t]
   \centering
   \begin{subfigure}[b]{0.5\textwidth}
       \centering
       \includegraphics[width=0.95\textwidth]{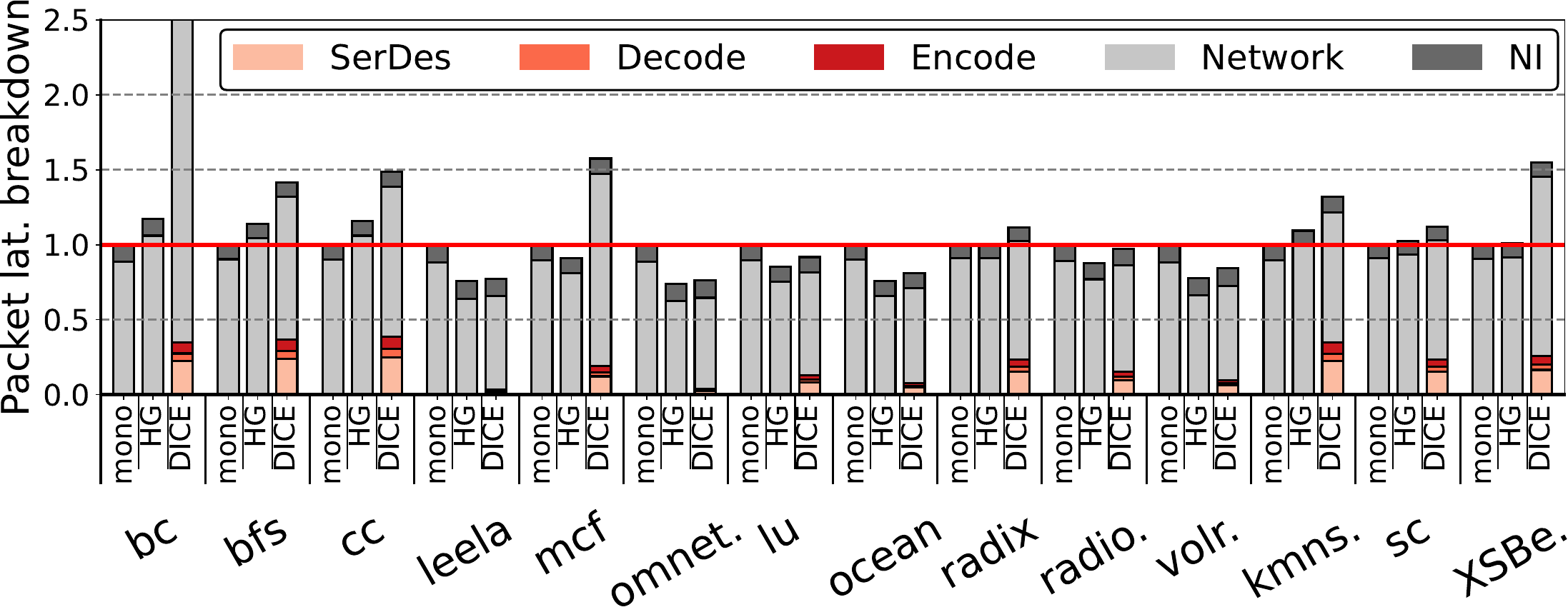}
       \caption{Normalized average network packet latency breakdown.}
       \label{fig:mon_vs_chip:a}
       \vspace*{0.2 cm}
   \end{subfigure} 
   \begin{subfigure}[b]{0.5\textwidth}
       \centering
       \includegraphics[width=0.95\textwidth]{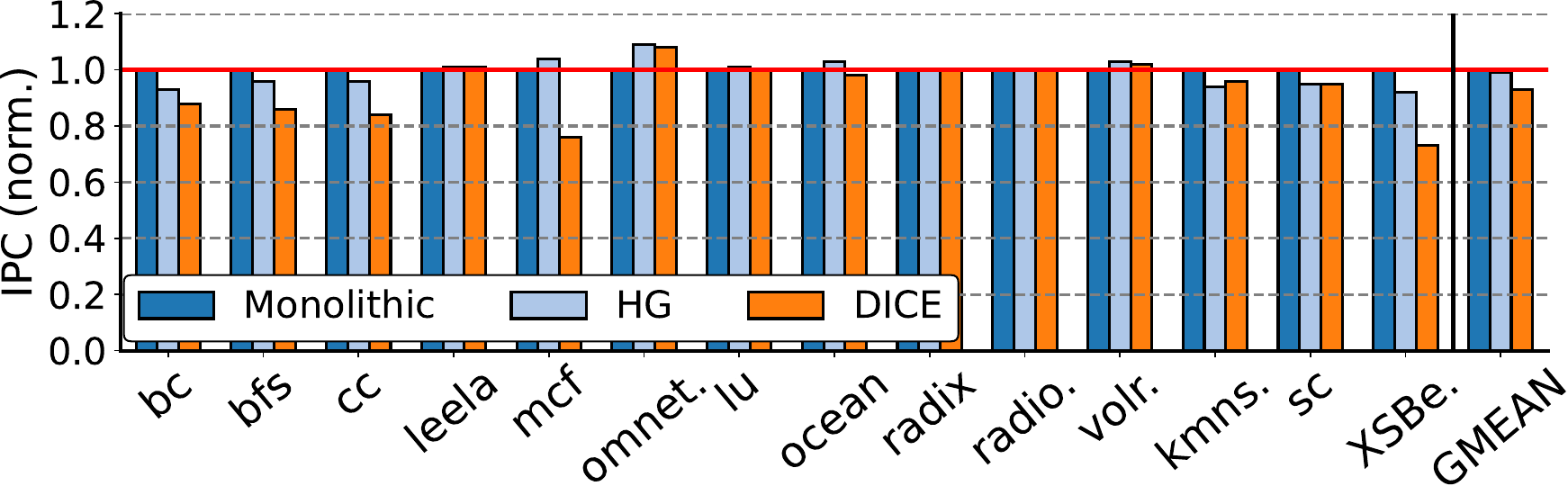}
       \caption{Normalized per-application IPC comparison.}
       \label{fig:mon_vs_chip:b}
   \end{subfigure}
   \caption{Impact of \phy{}-realistic modeling in \name{} on overall packet latency and system IPC.}
   \label{fig:mon_vs_chip}
   \tv{}
\end{figure}

% The end-to-end latency of an inter-chiplet transfer is thus runtime-determined:
% \[
% L_{\text{link}}(f_{\text{clk}}, W) =
% L_{\text{ser}}(f_{\text{clk}}) +
% L_{\text{tx/rx}} +
% L_{\text{ch}}\!\big(\mathrm{SNR}_{\text{base}}(t),\,\mathrm{jitter}(t)\big) +
% L_{\text{dec}}\!\big(N_{\text{iter}}\big) +
% L_{\text{rtx}}\!\big(P_{\text{fail}}(W,\,\mathrm{SNR}_{\text{base}})\big),
% \]
% where \(f_{\text{clk}}\) is the simulated frequency, \(W\) summarizes workload traits (e.g., LLC MPKI), \(N_{\text{iter}}\) is the LDPC iteration count chosen at runtime, and \(P_{\text{fail}}\) is the frame-error probability that triggers retransmission.

\textbf{Brief comparison of fixed-latency links and \name{}.} As shown in~\autoref{fig:mon_vs_chip}, we evaluate three cases: 1) a monolithic architecture with 32 cores connected via a 4$\times$8 mesh, denoted \textsf{Mono};
2) a chiplet architecture corresponding to~\autoref{fig:intro}, where inter-chiplet links are modeled using \textsf{HeteroGarnet}~\cite{kite} with fixed-latency, throttled channels, denoted \hg{}; and 3) our \phy{}-enabled chiplet architecture, \name{}. More details on simulation are in~\autoref{sec:metho}. From these figures, we draw 3 key observations.

\textit{First,} \name{} fundamentally changes the packet latency breakdown compared to both \textsf{Mono} and \hg{}, where average packet latency is largely dominated by network-interface (NI) queuing and network link traversal (Network). In \hg{}, the system is \textit{logically} chiplet-based, however, the latency profile mirrors \textsf{Mono} because detailed link and inter-chiplet \phy{} behaviors are not modeled. In contrast, in \name{}, a substantial fraction of end-to-end packet latency shifts to the chiplet \phy{} boundary, where it is spent on forward error correction (FEC) encoding, serialization/deserialization and modulation (SerDes), and FEC decoding and error correction (EC).

\textit{Second,} at the application level, we find that the IPC difference between \hg{} and \name{} is \emph{inconsistent} across workloads: \hg{} is sometimes optimistic and sometimes pessimistic, without a systematic bias that could be corrected by adjusting the throttling parameters (\autoref{sec:relevance}).

\textit{Third,} 
%for the benchmarks used in this study, \hg{} yields a geomean IPC that is effectively indistinguishable from the monolithic baseline, whereas \name{} yields a clear performance gap. This mismatch underscores that fixed-latency chiplet abstractions can mask critical \phy{}-induced costs and lead architects to misleading conclusions about chiplet-based systems.
%
for the collection of benchmarks that we chose as representative for this study (without prior knowledge of the results), \hg{} gives a geomean IPC that is practically indistinguishable from that of the monolithic baseline (\textsf{Mono}), whereas \name{} yields a clear performance gap.
This mismatch underscores that fixed-latency chiplet abstractions can mask critical \phy{}-induced costs and lead architects to misleading conclusions about chiplet-based systems (\autoref{sec:relevance}).

\textbf{Key Contributions.} We present \name{}, a gem5~\cite{gem5} module for in-simulation, end-to-end \phy{} modeling. Following the IEEE Heterogeneous Integration Roadmap~\cite{IEEE_HIR_HPC_2021}, \name{} models the major components of \phy{}-links, including:

\begin{itemize}[leftmargin=*, topsep=0pt, itemsep=1pt, parsep=0pt]

\item \textit{Channel noise.}
Inter-chiplet links are error-prone. We model inter-die links as additive white Gaussian noise (AWGN~\cite{proakis2007digital}) channels and fold clock jitter, inter-wire crosstalk, and channel operating conditions into the effective signal-to-noise ratio (SNR).

\item \textit{FEC encoding/decoding.}
We implement quasi-cyclic low-density parity-check (QC-LDPC~\cite{QC_LDPC}) forward error correction (FEC), a hardware-efficient scheme widely adopted in high-speed interconnects~\cite{LDPC_ssd, pcie6_fec_qa}. \name{} models both the QC-LDPC encoder and decoder. Further, because QC-LDPC decoding is NP-hard~\cite{LDPC}, decoders operate iteratively: they attempt to converge within a bounded iteration budget and, upon failure, trigger packet retransmission. This iterative behavior introduces inherent run-to-run latency variability. To reflect this accurately, \name{} calibrates decoder iteration budgets and per-iteration timing through hardware synthesis, yielding good latency models.

\item \textit{Serialization/Deserialization (SerDes) and signal modulation.}
After FEC encoding, die-edge SerDes performs parallel-to-serial (P2S) conversion, transmitting \textit{parallel} digits as high-speed \textit{serial} waveforms. \name{} models PAM-4 modulation---widely used in SerDes~\cite{IEEE_HIR_HPC_2021}---with waveform voltages and related parameters calibrated to public datasheets (\rebut{\autoref{tab:modeling_validation}}). 
At the receiver, PAM-4 demodulation reconstructs the digital bitstream, which then undergoes S2P conversion, and passed to the FEC decoder for error correction.

\item \textit{Router microarchitecture and inter-chiplet flow control.} We extend the router microarchitecture at the \phy{} boundary by integrating a dedicated \phy{}-level flow-control mechanism that combines FEC, modulation, and inter-chiplet retransmission support. This allows us to capture \phy{} serialization effects and FEC-induced backpressure in the end-to-end packet timing.

\end{itemize}

We compare \name{} against a actual AMD EPYC~9454P multicore processor by measuring core-to-core latency (\autoref{sec:vali}). We find that \name{} is closer to the measured C2C latencies of the actual 9454P than \textsf{HeteroGarnet}, providing a more faithful model of the underlying chiplet architecture.

% \item We perform an initial design space exploration with coherence as the only factor to showcase the differences that a chiplet design brings to the way we should perform evaluations henceforth. 

% \textit{5. Flexible design-space exploration (DSE).}
% \name{} supports CCD-private LLCs (AMD-style~\cite{Zen4_MICRO}) and globally shared LLCs across chiplets (Intel-style~\cite{Intel_sapphire}), as well as edge vs. central inter-chiplet router placements. System-wide studies with heterogeneous workloads reveal sizable deviations from monolithic baselines—especially under high traffic—where inter-chiplet latency and reduced effective LLC capacity amplify queuing and performance loss~\cite{chalmers_perf,MEMPLEX}.

\iffalse
{\color{purple}
\paragraph{Architectural lessons.}

performance deoends on temperature. increaing temperature now can affect performance which never happens before.

delta tempture -> delta temperature -> performance change. This is NEW!

For the first time temperature affects performance given that it does not trigger thermal urgency
}
\fi

\section{Inter-chiplet data communication and SOTA simulators for chiplet DSE}
\label{sec:relateworks}

\subsection{SOTA simulation frameworks for chiplet-based architectures}

Heterogeneous integration has emerged as a practical path to sustain performance scaling beyond the limits of Moore's Law. By assembling multiple smaller dies (chiplets) within a single package, designers can approximate large-die performance while improving yield and reducing cost. Unlike on-die communication, which remains purely digital over CMOS-driven wires, inter-chiplet links differ fundamentally~\cite{IEEE_HIR_HPC_2021}. CCD–IOD connections rely on high-speed SerDes signaling (\eg{}, NRZ or PAM4) driven by \phy{} circuits, where digital data are serialized, transmitted as modulated voltage waveforms, and then recovered through demodulation. 
As high-speed links continue to scale to higher data rates, \phy{}-links must operate under increasingly challenging channel conditions, including elevated jitter, crosstalk, insertion loss, and SNR-dependent bit errors. These impairments will soon necessitate the use of forward error correction (FEC) in next-generation \phy{} to ensure reliable inter-chiplet communication~\cite{IEEE_HIR_HPC_2021, 11150380}. Accordingly, accurate performance modeling of chiplet-based systems requires detailed \phy{}-level simulation that captures all of these considerations, rather than treating inter-die connections as ideal digital wires with a fixed delay.

A range of state-of-the-art (SOTA) simulators~\cite{gem5,garnet,Sniper,DARSim,booksim,Noxim,muchisim,BZSim,CNSim,RapidChiplet} have been proposed for chiplet-based multicore, each emphasizing different trade-offs in simulation detail and modeling focus. \textsf{gem5}~\cite{gem5} combined with \textsf{\rebut{HeteroGarnet}}~\cite{kite} offers full-system, cycle-accurate simulation of cores, caches, and routers, making it the de facto standard for microarchitectural studies. \textsf{Sniper}~\cite{Sniper} accelerates design space exploration via interval-based core models, providing faster but less detailed timing than \textsf{gem5}. \textsf{DARSIM}~\cite{DARSim} enables modular NoC evaluation using high-level in-order cores, while \textsf{BookSim~2.0}~\cite{booksim} and \textsf{Noxim}~\cite{Noxim} focus on lightweight, packet-level, trace-driven simulations that are ideal for router and traffic analysis but omit processor and memory hierarchy integration.

For larger-scale chiplet-oriented studies, \textsf{Muchisim}~\cite{muchisim} models distributed chiplet systems using simplified abstractions and software-managed coherence, whereas \textsf{BZSim}~\cite{BZSim} prioritizes simulation runtime via phase-based approximations suitable for workloads with modest network activity. \textsf{CNSim}~\cite{CNSim} introduces packet-parallel simulation to combine scalability with router-level timing fidelity. \textsf{RapidChiplet}~\cite{RapidChiplet} targets predicting inter-chiplet communication through high-level analytical models rather than cycle-accurate simulation.

Despite their strengths, none of these tools deliver fully realistic modeling of \phy{}-links (\eg{}, SerDes, FEC, channel noise) and their system-level impact (\eg{}, application runtime). To bridge this gap, \name{} extends \textsf{gem5} with a detailed \phy{}-level model that incorporates realistic cross-die communication, enabling end-to-end evaluation under true physical-layer constraints.

% \subsection{How serious are Ser/Der and Error}

% \red{Adding figures that show 1) the break down of average packet latency. How much time is spent on coding/decoding? how much is on error correction? draw this figure using different applications 2) draw another figure showinog how the coding/decoding latency percentage changes with different SNR\_{base} and modulation. What is the default SNR\_{base} we use now?}

\section{Design of \name{}}

We now detail the components of \name{}, starting with intra-CCD/IOD layout and on-die link configuration, followed by detailed inter-chiplet \phy{} modeling. We conclude with the router microarchitecture that connects multiple dies and implements inter-chiplet flow control.

\subsection{Intra-chiplet packet transmission.}

We modify \textsf{gem5}'s Python configuration to implement the system layout in~\autoref{fig:intro}, modeling an AMD EPYC–style architecture.

\textbf{CCD/IOD.}
Each CCD in \name{} integrates multiple cores with private L1/L2 caches and a shared LLC. Core-to-core communication within a CCD uses a CCD-local NoC over intra-die links. The default intra-CCD topology is a \m{2\times4} mesh with 8-core (aligned with EPYC ``Genoa''). \name{} models an IOD that aggregates memory controllers, DMA engines, and I/O. Placed in center (\autoref{fig:intro}), the IOD provides uniform access from CCDs to memory. Internally, we instantiate 4 \phy{} routers in the IOD in a \m{2\times2} mesh, with each \phy{} connected to 2 memory controllers (in total 8).

Reflecting the use of a less advanced process node for the IOD (\eg{}, 14\,nm \vs{} 5\,nm for CCDs), we model a higher NoC frequency for the CCDs than for the IOD. In \name{}, intra-CCD links run at 2.0\,GHz, are 128 bits wide, and incur 1-cycle router latency and 2-cycle link latency. In contrast, IOD links and routers operate at 1.0\,GHz, with the same 128-bit link width, 1-cycle router latency, and 2-cycle link transmission latency. In summary,~\autoref{tab:ccd_iod_config} lists the default parameters of the CCD and IOD models in \name{}.

\begin{table}[!h]
  \centering
  \scriptsize
  \caption{Default for intra-CCD/IOD NoC.}
  \begin{tabular}{lccccc}
    \toprule
    \textbf{} & \textbf{Topology} & \textbf{Link width} & \textbf{Freq} & \textbf{Router lat} & \textbf{Link lat} \\
    \midrule
    CCD & mesh \(2\times4\)& 128-bit & 2.0\,GHz & 1 cycle & 2 cycles\\
    IOD & mesh \(2\times2\) & 128-bit & 1.0\,GHz & 1 cycle & 2 cycles\\
    \bottomrule
  \end{tabular}
  \label{tab:ccd_iod_config}
\end{table}

\subsection{Inter-chiplet packet transmission.}
\label{sec:rw}

%\begin{figure}[!t]
%  \begin{center}
%    \includegraphics[width=0.475\textwidth]{figs/text/yuan/comm.pdf}
%  \end{center}
%  \caption{Inter-chiplet communication modeling in \textsf{HeteroGarnet} in \textsf{gem5}, throttling link bandwidth to simulate SerDes.}
%  \label{fig:comm}
%\end{figure}

\begin{figure}[!t]
  \begin{center}
    \includegraphics[width=0.475\textwidth]{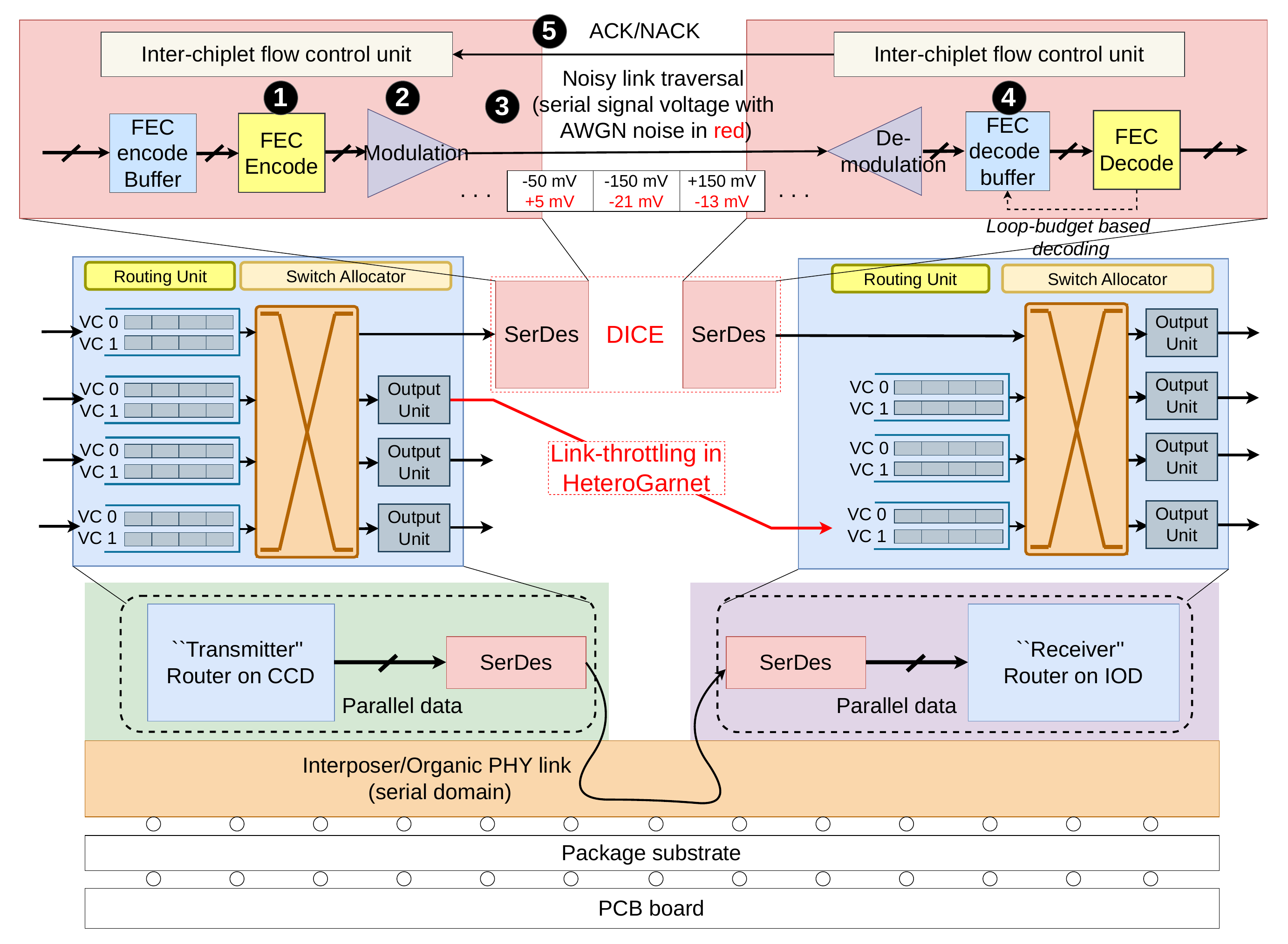}
  \end{center}
  \caption{Detailed end-to-end inter-chiplet communication modeling in \name{}, capturing FEC en-/de-coding, modulation/de-modulation, noise injection, and inter-chiplet flow control.}
  \label{fig:serdes}
  \tv{}
\end{figure}

Conventional simulators approximate inter-chiplet communication with limited fidelity (\autoref{sec:relateworks}). For example, \textsf{gem5 HeteroGarnet}~\cite{kite} emulates SerDes by throttling bandwidth between inter-chiplet routers (\autoref{fig:serdes}---\textsf{\rebut{HeteroGarnet}}) rather than modeling the \phy{}. To capture actuate cross-die packet transmission, \name{} models the full inter-chiplet transmission stack (\autoref{fig:serdes}---\name{}), including \ding{202} FEC encoding, \ding{203} modulation, \ding{204} noisy link traversal, \ding{205} demodulation, and FEC-decoding---with \ding{206} integrated inter-chiplet flow control to manage backpressure and retransmissions. Next, we detail each of the components.

%%%%%%%%%%%%%%%%%%%%%%%%%%%%%%%%%%%%%%%%

\subsection{Forward error correction (FEC) encoding} 
\label{sec:fec:encode}

Unlike CRC-only schemes that trigger retransmission upon error detection~\cite{koopman2002crc}, \name{} employs forward error correction (FEC)~\cite{ru2008}, which proactively corrects bit errors at the receiver and significantly reduces replays. In \name{}, the transmit-side \phy{} router aggregates flits, applies FEC encoding, and then serializes them for transmission.

\textbf{QC-LDPC encoding.}
\name{} employs Quasi-Cyclic Low-Density Parity-Check (QC-LDPC) codes~\cite{QC_LDPC}, a hardware-efficient FEC scheme extensively used in SSDs~\cite{ssd_LDPC_ACM,LDPC_ssd,LDPC}. A QC-LDPC code is specified by a sparse parity-check matrix \m{H\in\{0,1\}^{m\times n}}. A binary vector \m{\mathbf{c}\in\{0,1\}^{n}} is a valid codeword iff:
\begin{equation}
    H\mathbf{c}^T \equiv \mathbf{0}\pmod{2}.
\end{equation}
Here, \m{n=k+m} is the codeword length,
\m{k} is the number of bits in the packet-under-transmission, and 
\m{m} is the number of parity bits (rows of \m{H}).
The code rate \m{R} is defined as:
\begin{equation}
    R=\frac{k}{n}=1-\frac{m}{n},
\end{equation}
which shows the fraction of the codeword devoted to the original message.

\begin{figure}[t]
  \centering
  \begin{subfigure}{\linewidth}
    \centering
    \includegraphics[width=.9\linewidth]{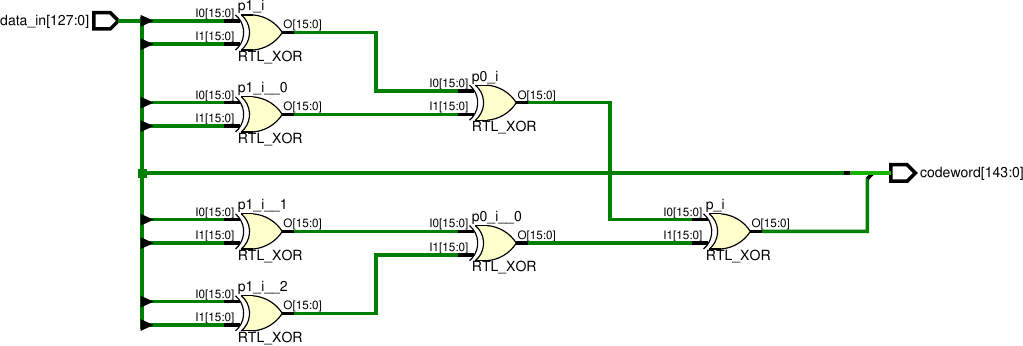}
    \caption{Flit-level (128-bit) FEC encoder.}
    \label{fig:fec-en-hw:128}
  \end{subfigure} \\
  \begin{subfigure}{\linewidth}
    \centering
    \includegraphics[width=.3\linewidth, angle=-90]{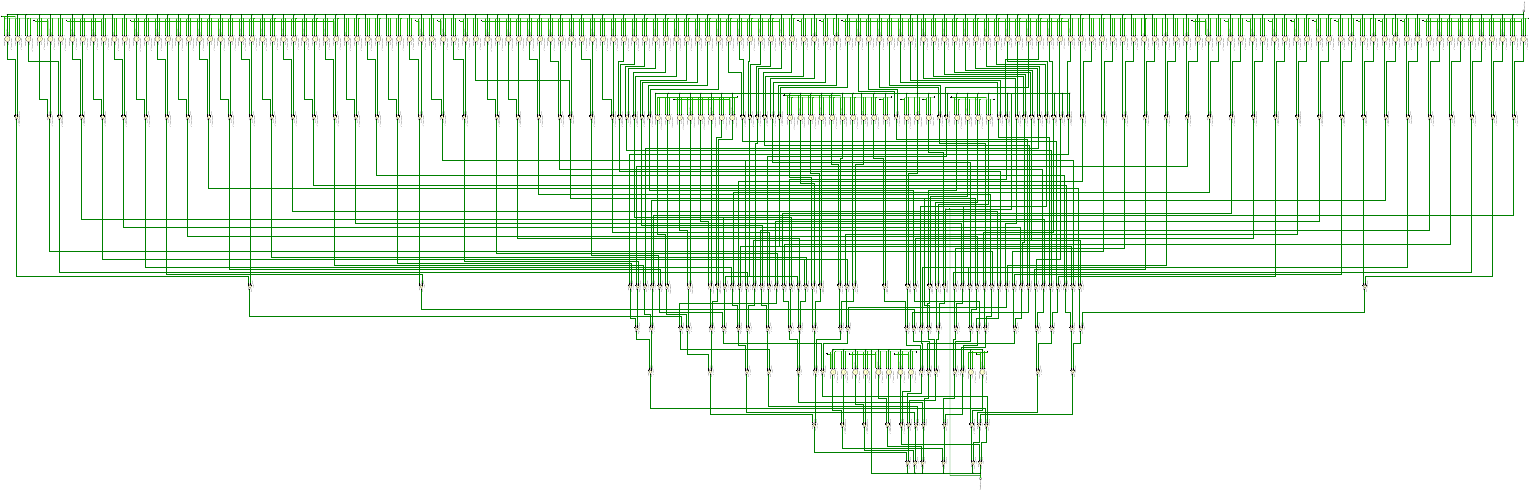}
    \caption{Packet-level (64-byte) FEC encoder.}
    \label{fig:fec-en-hw:768}
  \end{subfigure}
  \caption{Logic synthesis results for flit- and packet-level FEC encoders. Layouts visualized in Xilinx Vivado for clarity.
  \review{B.Other\\E.Other}
  \rebut{Takeaway: Flit-level FEC can be implemented with a simple three-level XOR-net, whereas packet-level FEC significantly increases hardware complexity. 
  }
  }
  \label{fig:fec-en-hw}
  \tv{}
\end{figure}

\textbf{FEC en/decoding granularity.}
\name{} performs FEC at \emph{flit} granularity: \dc{each 128-bit flit} is encoded and decoded independently using a parity-check matrix \m{H} shared among \phy{} sender-receiver pairs, ensuring consistent encode/decode/error correction across chiplets. Control packets consist of a single \textsf{HEAD-TAIL} flit, while data packets consist of 6 flits---1 \textsf{HEAD}, 4 \textsf{BODY} flits carrying a \m{64\,\mathrm{Byte}} cache line, and 1 \textsf{TAIL} flit. This flit-level granularity is chosen because: 1) it keeps the encoder’s hardware footprint small and timing-friendly (see~\autoref{fig:fec-en-hw} for FEC-encoder hardware), and 2) it enables more flexible inter-chiplet flow control (\autoref{sec:router-micro}).

% \begin{figure}[!t]
%   \centering
%   \includegraphics[width=0.5\textwidth]{figs/text/yuan/ldpc_pam4/ldpc_pam4_15.0.pdf}
%   \caption{Pre-/Post-FEC and error corrected flits within a noisy channel.}
%   \label{fig:waterfall_15}
% \end{figure}

\begin{figure}[!t]
  \centering
  %----- Subfigure 1 -----
  \begin{subfigure}{0.45\textwidth}
    \centering
    \includegraphics[width=\textwidth]{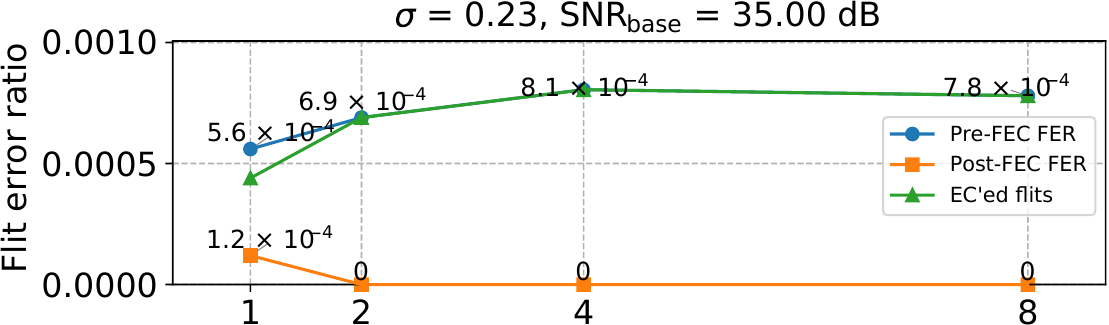}
    \label{fig:waterfall_35}
    \vspace*{-0.25 cm}
  \end{subfigure}
  %----- Subfigure 2 -----
  \begin{subfigure}{0.45\textwidth}
    \centering
    \includegraphics[width=\textwidth]{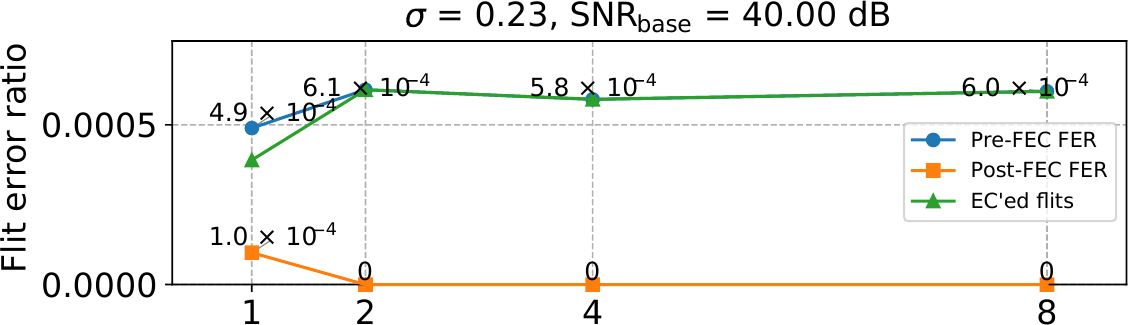}
    \label{fig:waterfall_400}
    \vspace*{-0.25 cm}
  \end{subfigure}
  %----- Subfigure 3 -----
  \begin{subfigure}{0.45\textwidth}
    \centering
    \includegraphics[width=\textwidth]{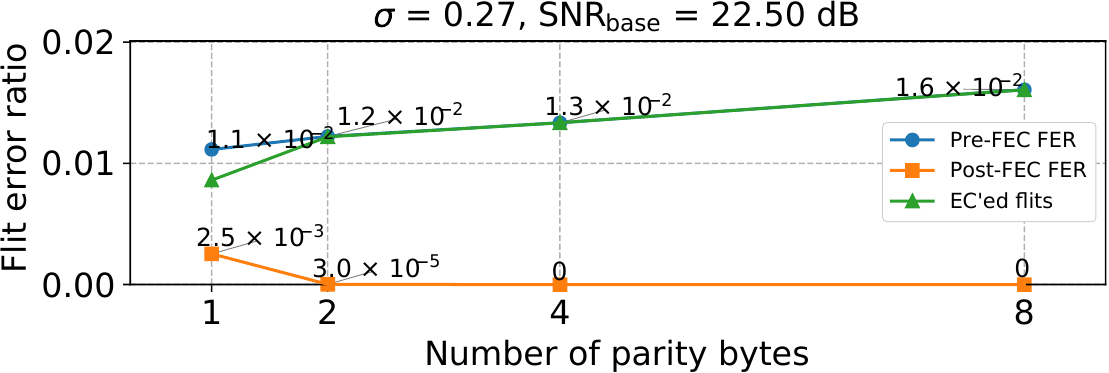}
    \label{fig:waterfall_225}
    \vspace*{-0.25 cm}
  \end{subfigure}
  \caption{
  \review{E.Other}
  \rebut{Pre- and post-FEC FER and number of error-corrected flits under three baseline SNRs with varying parity-byte configurations. Takeaway: A 2-byte parity per flit %strikes 
  sits at a sweet spot between overhead and Post-FEC FER.}
  }
  \label{fig:waterfall_all}
  \tv{}
\end{figure}

\textbf{Sensitivity study on code rate $\mathbf{R}$.}
We first select the code rate \m{R}. Unlike BCH/Hamming, QC\text{-}LDPC provides no bounded-distance guarantee, making \m{R} a design trade-off: lower \m{R} (more parity) strengthens error correction (EC) but consumes more inter-chiplet bandwidth; higher \m{R} reduces overhead but weakens correction. We choose \m{R} using the sensitivity study in \autoref{fig:waterfall_all}, which reports both the pre-FEC flit error ratio (FER) and the post-FEC FER under a Gaussian-noise channel at a representative signal-to-noise ratio of \(\mathrm{SNR}_{\text{base}}=\dc{35.0\,\mathrm{dB}}\). Additional sensitivity experiments on \(\mathrm{SNR}_{\text{base}}\) are discussed in \autoref{sec:ber}. All results are decoded using a 4-iteration loop-budget layered Min-Sum FEC decoder (\autoref{sec:fec:decode}).

% \begin{figure}[!t]
%   \centering
%   \includegraphics[width=0.45\textwidth]{figs/text/yuan/ldpc_pam4/ldpc_pam4_35.0.pdf}
%   \caption{Pre-/Post-FEC and error corrected flits within a typical noise channel (\m{\mathrm{SNR_{base}=\dc{35.0\,dB}}}).}
%   \label{fig:waterfall_35}
%   \vspace*{-0.25 cm}
% \end{figure}

% \begin{figure}[!t]
%   \centering
%   \includegraphics[width=0.45\textwidth]{figs/text/yuan/ldpc_pam4/ldpc_pam4_40.0.pdf}
%   \caption{Pre-/Post-FEC and error corrected flits within a typical noise channel (\m{\mathrm{SNR_{base}=\dc{40.0\,dB}}}).}
%   \label{fig:waterfall_400}
% \end{figure}

% \begin{figure}[!t]
%   \centering
%   \includegraphics[width=0.45\textwidth]{figs/text/yuan/ldpc_pam4/ldpc_pam4_22.5.pdf}
%   \caption{Pre-/Post-FEC and error corrected flits within a typical noise channel (\m{\mathrm{SNR_{base}=\dc{22.5\,dB}}}).}
%   \label{fig:waterfall_225}
%   \tv{}
% \end{figure}

From the figure we observe a sweet spot at 2 parity bytes per flit (\m{R \approx \dc{0.88}}): higher \m{R} under-provisions parity and increases post-FEC error rates, whereas lower \m{R} yields diminishing returns in correction strength. Consequently, \name{} adopts \m{R \approx 0.88} by default---1 parity bit per flit-byte (16 parity bits for a 128-bit flit)---balancing post-FEC reliability and inter-chiplet bandwidth efficiency.

As a comparison to \dc{\m{\mathrm{SNR_{base}=\dc{35.0\,dB}}}}, \autoref{fig:waterfall_all} also shows results at \dc{\m{\mathrm{SNR_{base}=\dc{40.0\,dB}}}} and \dc{\m{22.5\,dB}}. 
First, \dc{\m{\mathrm{SNR_{base}=\dc{40.0\,dB}}}} achieves results similar to \dc{\m{\mathrm{SNR_{base}=\dc{35.0\,dB}}}}, which indicates that once \dc{\m{\mathrm{SNR_{base}}}} exceeds \dc{\m{\mathrm{35.0\,dB}}}, the channel noise is dominated by crosstalk and jitter, and a cleaner baseline SNR does not provide much additional benefit. 
Second, in contrast, under noisier channel conditions, 2 parity bytes are no longer sufficient to maintain a satisfactory post-FEC FER. In this regime, there are 3 possible trade-offs: 1) increase the number of parity bytes (\eg{}, from 2 to 4, as shown in the figure) to better tolerate channel noise, at the cost of reduced effective data bandwidth; 2) maintain a 2-byte parity but retransmit flits that cannot be corrected, preserving raw channel bandwidth but increasing packet turnaround time and tax application runtime; or 3) increase the FEC decoder iteration budget to improve error-correction capability, at the expense of additional hardware complexity and power.

%%%%% Block-circulant construction

\textbf{Parity function construction.} 
% \red{Stefanos: I already fixed this}
To generate parity bits, \name{} constructs a parity-check matrix \m{H} for the LDPC code. The process begins with a compact 
\emph{base matrix} \m{B \in \mathbb{Z}^{m_b \times n_b}}, where
$m = m_b Z, n = n_b Z$. The expansion factor \m{Z} controls hardware parallelism: a larger \m{Z} increases parallelism.
%Note that the \emph{code rate} ratio 
%\[
%R = 1 - \frac{m_b}{n_b} = 1-\frac{m}{n}
%\]
%defines the , i.e., the fraction of bits devoted to the original data.
Each element \m{b_{i,j}} in \m{B} represents either:
%\begin{itemize}
%    \item 
\m{-1}, denoting an all-zero block, or
%    \item 
a non-negative shift value \m{s \in \{0, 1, \ldots, Z-1\}}, denoting a right-rotated identity matrix.
%\end{itemize}
%
Each shift value \m{s} corresponds to a \m{Z \times Z} \emph{circulant permutation matrix}:
\[
P(s) = \text{RotateRight}_Z(I_Z, s), \quad P(-1) = 0_{Z \times Z},
\]
where \m{I_Z} is the \m{Z \times Z} identity matrix and 
\m{\text{RotateRight}_Z(I_Z, s)} performs a cyclic right shift of each row by \m{s} positions.
Replacing every entry \m{b_{i,j}} in \m{B} with its corresponding block \m{P(b_{i,j})} 
produces the full parity-check matrix:
\[
H = [\, P(b_{i,j}) \,]_{i \in [0, m_b - 1],\, j \in [0, n_b - 1]}, 
\quad m = m_b Z, \quad n = n_b Z.
\]

This \emph{block-circulant} construction yields a highly regular and hardware-friendly matrix, 
since each \m{P(s)} can be implemented using simple shift registers or address rotations, 
avoiding the complex interconnects of unstructured LDPC codes.

\textbf{Example.}
For \m{Z = 16}, each \m{P(s)} is a \m{16 \times 16} rotation of the identity matrix.
A 128-bit flit is divided into 8 16-bit chunks \m{\{u_0, u_1, \ldots, u_7\}}. The parity block \m{p} is computed as:
%\[
%p = P(0)u_0 \oplus P(3)u_1 \oplus P(7)u_2 \oplus P(11)u_3 
%   \oplus P(2)u_4 \oplus P(9)u_5 \oplus P(14)u_6 \oplus P(5)u_7.
%\]
\begin{equation}
\label{eq:fec}
\begin{aligned}
\mathbf p
&= P(0)\,\mathbf u_0 \;\oplus\; P(3)\,\mathbf u_1 \;\oplus\; P(7)\,\mathbf u_2 \;\oplus\; P(11)\,\mathbf u_3 \; \oplus \\\
&\quad \; P(2)\,\mathbf u_4 \;\oplus\; P(9)\,\mathbf u_5 \;\oplus\; P(14)\,\mathbf u_6 \;\oplus\; P(5)\,\mathbf u_7.
\end{aligned}
\end{equation}
The final codeword concatenates data and parity bits:
\[
c = [\, u_0 \| u_1 \| \cdots \| u_7 \| p \,].
\]

%This structured, block-circulant LDPC design enables efficient encoding and decoding with predictable memory access and high hardware efficiency.

\textbf{FEC-encoder latency.}
We implement the QC-LDPC encoder described in~\autoref{eq:fec} in Verilog. We use \dc{Yosys}\footnote{\url{https://github.com/YosysHQ/yosys}} for synthesis and \dc{OpenSTA}\footnote{\url{https://github.com/The-OpenROAD-Project/OpenSTA}} for static timing analysis, both targeting the \dc{TSMC 40\,nm} standard-cell library. Our analysis shows that FEC granularity has a significant impact on hardware cost and timing. With a 128-bit input (flit-level), the encoder maps to \dc{7} 16-bit XOR gates using \dc{175} standard cells (\autoref{fig:fec-en-hw:128}) and meets the target 2.0\,GHz clock constraint. Multipliers are optimized away since all coefficients in \m{H} are constants. In contrast, scaling the input width to 768 bits (packet-level, corresponding to a full data packet) increases logic complexity to \dc{2320} cells (\autoref{fig:fec-en-hw:768}) and fails to meet the 2.0\,GHz timing target. These results motivate the choice of \emph{flit}-level FEC in \name{}, ensuring the encoder remains off the router pipeline’s critical path in both CCD and IOD.

\begin{figure}[!t]
  \centering
  \includegraphics[width=0.45\textwidth]{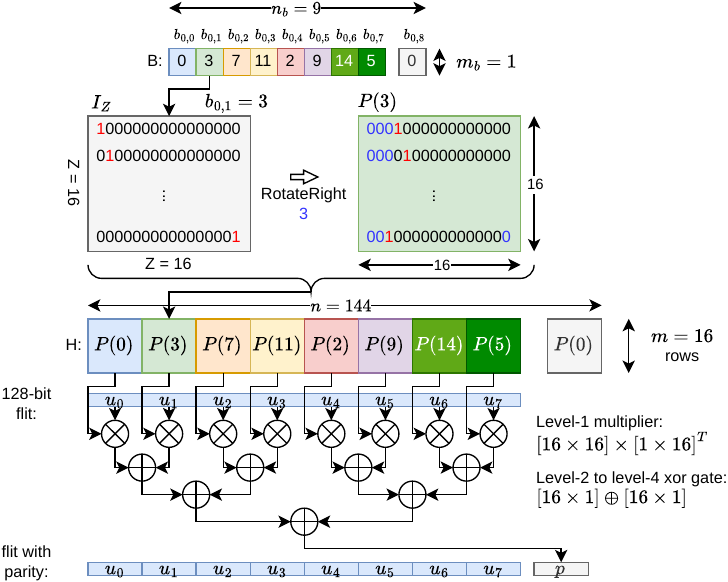}
  \caption{FEC-encoding for 128-bit flit with 2-byte parity bits.}
  \label{fig:fec-en}
\end{figure}

%%%%%%%%%%%%%%%%%%%%%%%%%%%%%%%%%%%%%%%%5

%\subsection{Modulation and channel noise modeling}
\subsection{Modulation}
\label{sec:ber}

After FEC encoding, digital signals are modulated and transmitted over the inter-chiplet channel.

\textbf{PAM4 modulation.}
\name{} employs PAM4~\cite{stauffer2018serdes} with Gray mapping, where 2-bit symbols \m{[00,01,11,10]} correspond to amplitudes \m{[-3d,-d,+d,+3d]}. In \name{}, we use an interposer-level swing of \m{[-150,-50,+50,+150]\,\mathrm{mV}}, \ie{}, \m{d=50\,\mathrm{mV}}, representative of short-reach chiplet links on silicon interposers. This low-swing operation significantly reduces I/O power and is consistent with modern die-to-die PAM4 \phy{} implementations~\cite{IEEE_HIR_HPC_2021}.

\textbf{Example: PAM4.} Transmitting the byte word \m{\dc{[01,00,10,10]}} produces symbol pairs \m{[01,00,10,10]}, which map to amplitudes \m{[-d,-3d,+3d,+3d]}. With \m{d=50\mathrm{mV}}, the resulting serial waveform is \m{x=[-50,-150,+150,+150]\,\mathrm{mV}}.

\subsection{Channel noise modeling}
\label{sec:awgn}

After modulation symbols are transmitted over the inter-chiplet channel where \emph{noise} (in the form of channel, jitter, and cross talk noise) is added.

\textbf{AWGN channel noise.}
\name{} models the channel noise as additive white Gaussian noise (AWGN), accounting for two dominant high-speed link impairments---\textit{1) jitter}~\cite{TI} and \textit{2) crosstalk (XT)}~\cite{xtalk_tsv}. Together with the channel's intrinsic operating conditions, these impairments determine link quality, which we quantify using signal-noise ratio (SNR)~\cite{BER_2,BER_3}.

\textbf{Jitter.}
Jitter refers to random variations in signal timing that translate clock-edge uncertainty into waveform errors (samples too early or late), thereby reducing SNR, closing the eye diagram, and degrading link quality. A widely used jitter model~\cite{CDR,equ,modulation_jitter} is: 
\begin{equation}
\small
\label{eq:snr_jitter} 
\mathrm{SNR}_{\mathrm{jitter,linear}} 
\approx 
\left(
    \frac{T_{\mathrm{sym}}}{\pi \sigma_t}
\right)^{2} 
\;\Rightarrow\; 
\mathrm{SNR}_{\mathrm{jitter,dB}} 
\approx 
20 \log_{10}
\left(
    \frac{T_{\mathrm{sym}}}{\pi \sigma_t}
\right), 
\end{equation}
where $\sigma_t$ denotes the root-mean-square (RMS) clock-edge timing error and $T_{\mathrm{sym}}$ the symbol period.
In \name{}, we set $T_{\mathrm{sym}}$ according to the network clock rate (32 Gb/s, following AMD's Infinity Fabric) and assume $\sigma_t \approx 1\,\mathrm{ps}$~\cite{xtalk_db, pcie5_qa}, yielding an average $\mathrm{SNR}_{\mathrm{jitter,dB}} \approx \dc{26.0}\,\mathrm{dB}$.

\textbf{Crosstalk (XT).}
XT is hardware-induced coupling from aggressor lanes into a victim channel; its severity depends on interconnect geometry (\eg{}, wire spacing and length)~\cite{distance} and the modulation format~\cite{BER_3,SerDes1}. By default, \name{} sets \m{\mathrm{SNR}_{\mathrm{XT,dB}}\approx \dc{20.0}\,\mathrm{dB}} guided by public data~\cite{UCIeSignalIntegrity}.

\textbf{Calculating effective SNR and BER.}
We quantify the end-to-end link quality using an \emph{effective} SNR that aggregates independent impairment sources. Let $\mathrm{SNR}_{\mathrm{base}}$ represent the intrinsic channel noise determined by its physical characteristics and operating conditions. Converting all SNR terms into \emph{linear} scale, the effective SNR is obtained using the harmonic-sum rule~\cite{proakis2007digital}:
\begin{equation}
\small
\label{eq:snr_eff}
\frac{1}{\mathrm{SNR}_{\mathrm{eff,linear}}}
=
\frac{1}{\mathrm{SNR}_{\mathrm{base,linear}}}
+
\frac{1}{\mathrm{SNR}_{\mathrm{jitter,linear}}}
+
\frac{1}{\mathrm{SNR}_{\mathrm{XT,linear}}}.
\end{equation}

% \begin{table}[!t]
% \centering
% \footnotesize
% \caption{xxx}
% \label{tab:bertime}
% \begin{tabular}{M{1.5cm}|M{1.5cm}|M{3cm}}
% \hline
% \textbf{Order of BER} & \textbf{Bits before 1 error} & 
% \textbf{Time to first error 256Gb/s (128-bit @ 2.0GHz)} \\
% \hline
% \m{10^{-2}}  & 1.00$\times$\m{10^{2}}  & 0.391 $n$s \\
% \m{10^{-5}}  & 1.00$\times$\m{10^{5}}  & 0.391 $\mu$s \\
% \m{10^{-8}}  & 1.00$\times$\m{10^{8}}  & 0.391 $m$s \\
% \m{10^{-12}} & 1.00$\times$\m{10^{12}} & 3.91 s \\
% \hline
% \end{tabular}
% \end{table}

\begin{lstlisting}[
    float,
    caption={Error injector pseudocode},
    label={code:error-inject},
    floatplacement=t, 
    language=c++,
    numbers=left,
    numberstyle=\tiny\color{gray},
    numbersep=5pt,
    xleftmargin=1.5em,
    basicstyle=\sffamily\footnotesize,
    escapeinside={(@}{@)}
  ]
/*Initialize random generator with SNR_eff,
   which is determined by (@\textbf{jitter, XT, SNR\_base}@)*/
(@\textbf{sigma <- Es / SNR\_eff\_linear}@)
rng <- NormalDist(mean=0, stddev=sigma)
for each flit in network:
  msg_byte <- flit.get_msg_byte()
  for each 2-bit PAM4 symbol in msg_byte:
    // {00->-3d, 01->-d, 11->+d, 10->+3d}
    x <- gray_map(bits) // From bits to symbols       
    n <- rng()          // AWGN noise sample
    (@\textbf{y <- x + n}@)             // Adding noise to symbol
    NetworkLink.send(y) // Signal sent via link
\end{lstlisting}

\textbf{Error injection.}
Driven by \m{\mathrm{SNR}_{\mathrm{eff}}}, \name{} implements an \emph{error injector} in gem5 (\autoref{code:error-inject}) that corrupts transmitted PAM4 symbols with AWGN.
As shown in~\autoref{code:error-inject}, with Gray mapping and
\m{\mathcal{X}=\{-3d,-d,+d,+3d\}}
at \m{d=50\,\mathrm{mV}}~\cite{SerDes1, 11150380},
a symbol \m{x\in\mathcal{X}} is transmitted as:
\[
  y = x + n,\qquad n \sim \mathcal{N}(0,\sigma^2),
\]
where \m{\sigma^2} is the voltage noise variance implied by \m{\mathrm{SNR}_{\mathrm{eff}}}. With average energy per symbol \m{E_s=\frac{1}{4}\sum_{x\in\{-3d,-1d,1d,3d\}}x^2=5d^2}, \m{\sigma} is given by: 
\[
\small
\sigma^2 = \frac{E_s}{\text{SNR}_{\text{eff,linear}}}
         = \frac{5d^2}{\text{SNR}_{\text{eff,linear}}}.
\]
Following UCIe practice for short-reach inter-chiplet links, we assume a nominal \m{\mathrm{SNR}_{\mathrm{base,dB}} \approx \dc{35.0}\,\mathrm{dB}}~\cite{IEEE_HIR_HPC_2021,UCIe_HotChips23}.
After incorporating jitter and crosstalk impairments, this corresponds to an effective link quality of \m{\mathrm{SNR}_{\mathrm{eff,dB}} \approx \dc{19.0}\,\mathrm{dB}}, which yields a baseline noise variance of \m{\sigma \approx \dc{12.7}\,mV}. Lastly, beyond the defaults, \name{} exposes all knobs for jitter level, crosstalk strength, and baseline SNR at run time to enable flexible DSE.

\textbf{Example: corrupting transmitted symbols.}  
Consider the PAM4 sequence 
\m{x=[-50,\,-150,\,+150,\,+150]\,\mathrm{mV}} 
with a noise deviation of 
\m{\sigma \approx \dc{12.7}\,\mathrm{mV}}.  
For a sample noise realization 
\m{n=[+5.0,\,-21.0,\,-13.0,\,+8.0]\,\mathrm{mV}}, 
the transmitted waveform becomes
\m{
  y = x + n = \underline{[\dc{-45.0},\,\dc{-171.0},\,\dc{+137.0},\,\dc{+158.0}]\,\mathrm{mV}.}
}

% \begin{figure}[!t]
%   \centering
%   \includegraphics[width=0.2\textwidth]{figs/text/yuan/code/snr_base/pam4_normal_15.0-slim.pdf}
%   \caption{\m{\mathrm{SNR_{base}}=\dc{15.0}\,\mathrm{dB}}.}
%   \label{fig:snr_base:15}
%   \tv{}
% \end{figure}

% \begin{figure}[!t]
%   \centering
%   \includegraphics[width=0.2\textwidth]{figs/text/yuan/code/snr_base/pam4_normal_35.0-slim.pdf}
%   \caption{\m{\mathrm{SNR_{base}}=\dc{35.0}\,\mathrm{dB}}.}
%   \label{fig:snr_base:35}
%   \tv{}
% \end{figure}

\begin{figure}[!t]
  \centering
  \begin{subfigure}{0.24\textwidth}
    \centering
    \includegraphics[width=\textwidth]{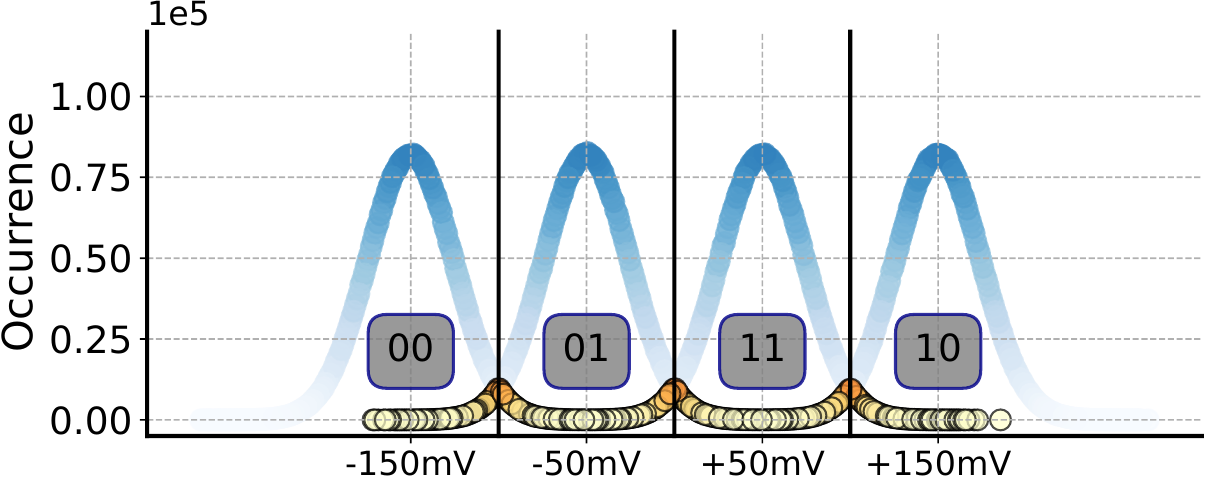}
    \caption{\m{\mathrm{SNR_{base}} = 15.0\,\mathrm{dB}}}
    \label{fig:snr_base:15}
  \end{subfigure}
  \begin{subfigure}{0.24\textwidth}
    \centering
    \includegraphics[width=1.08\textwidth]{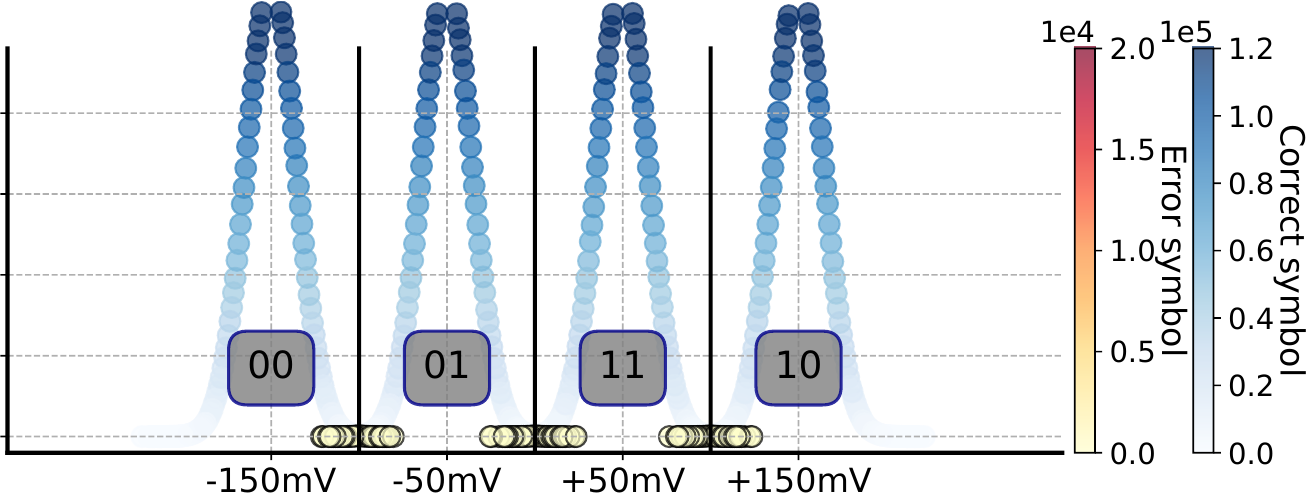}
    \caption{\m{\mathrm{SNR_{base}} = 35.0\,\mathrm{dB}}}
    \label{fig:snr_base:35}
  \end{subfigure}
  \caption{
  \review{E.Other}
  Error distribution under two example base \m{\mathrm{SNRs}}.
  \rebut{Takeaway: Higher SNR incurs fewer errors.}
  }
  \label{fig:snr_base:all}
\end{figure}

\begin{figure}[!t]
  \centering
  \includegraphics[width=0.45\textwidth]{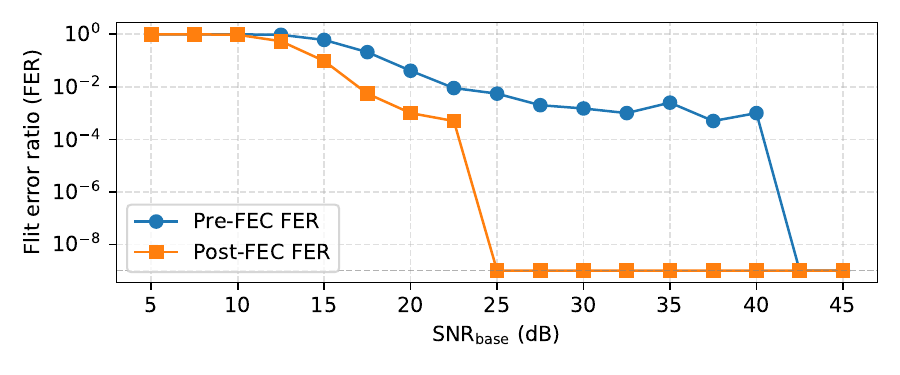}
  \caption{
  Pre- and post-FEC FER sensitivity across varying base \m{\mathrm{SNR}} values.
  \review{E.Other}
  \rebut{Takeaway: when SNR falls below 25\,dB (\ie{}, in a highly noisy channel), 2-byte parity FEC becomes unreliable, necessitating either additional parity bytes or higher-level error detection mechanisms such as CRC. This paper adopts 35\,dB as the baseline SNR based on publicly available data~\cite{IEEE_HIR_HPC_2021}.}
  }
  \label{fig:pre-fec:modu}
  \tv{}
\end{figure}

In \autoref{fig:snr_base:15} and \autoref{fig:snr_base:35}, we visualize PAM4 symbol distributions under AWGN for \m{\mathrm{SNR}_{\mathrm{base,dB}}}=15.0 and 35.0, respectively. An \textit{error symbol} is counted when noise drives a symbol across an adjacent voltage region (\eg{}, -150\,mV$\rightarrow$-50\,mV). We further report the pre-FEC FERs across a span of \m{\mathrm{SNR}_{\mathrm{base,dB}}} in \autoref{fig:pre-fec:modu}.

As shown in \autoref{fig:snr_base:15} and \autoref{fig:snr_base:35}, increasing $\mathrm{SNR}_{\mathrm{base,dB}}$---reflecting a cleaner, better-manufactured channel---reduces the number of error symbols. Further, with 2 parity bytes, \autoref{fig:pre-fec:modu} shows that when the channel SNR falls below \dc{25.0\,$\mathrm{dB}$}, a growing fraction of errors become uncorrectable by FEC and must instead be handled at a higher protocol level, such as retransmission via CRC~\cite{UCIe_BER}. 
% Further discussion of these trade-offs is in \autoref{sec:eva-1}.

\subsection{De-modulation}
\label{sec:llr}

On the other side of the inter-chiplet channel, the receiver performs FEC decoding (\autoref{sec:fec:decode}) and error correction~\cite{ru2008}, reconstructs flits, and forwards them to the downstream router. The first step, however, is to demodulate the received signal by computing \emph{channel LLRs} or \emph{log-likelihood ratios}, which represent a ``soft'' confidence for the received symbol bits. The reason LLRs are important is because they are the input to the decoder described in \autoref{sec:fec:decode}.

At the receiver, \name{} computes bit-wise 
%log-likelihood ratios (LLRs) 
LLRs for each received PAM4 symbol (encoding 2 bits). 
Let \m{L_{\mathrm{ch},k}} denote the LLR for bit \m{b_k}. Its magnitude \m{|L_{\mathrm{ch},k}|} captures confidence (\eg{}, \m{|L_{\mathrm{ch},k}|=3.0} is high; \m{|L_{\mathrm{ch},k}|=0.2} is low), while its sign encodes polarity (\m{L_{\mathrm{ch},k}>0}$\Rightarrow$\texttt{0}, \m{L_{\mathrm{ch},k}<0}$\Rightarrow$\texttt{1}). For PAM4 with Gray-mapping \m{[00,01,11,10] \Rightarrow [-3d,-d,+d,+3d]}, the symbol-bit subsets are:

{\scriptsize
\[
\mathcal{X}_{\text{MSB}}^{(1)}=\{+d,+3d\},
\mathcal{X}_{\text{MSB}}^{(0)}=\{-3d,-d\},
\mathcal{X}_{\text{LSB}}^{(1)}=\{-d,+d\},
\mathcal{X}_{\text{LSB}}^{(0)}=\{-3d,+3d\}.
\]
}

Given a received symbol \m{y}, the LLR \m{L_{\text{ch}}} for bit \m{b_k} is~\cite{669123}:
{
\scriptsize
\[
L_{\text{ch}}(b_k)
= \log 
\frac{\sum\limits_{x\in\mathcal{X}_k^{(0)}} 
\exp\!\big(-\tfrac{(y-x)^2}{2\sigma^2}\big)}
{\sum\limits_{x\in\mathcal{X}_k^{(1)}} 
\exp\!\big(-\tfrac{(y-x)^2}{2\sigma^2}\big)}
\;\approx\;
\frac{1}{2\sigma^2}\!\left(
\min_{x\in\mathcal{X}_k^{(1)}}(y-x)^2
-
\min_{x\in\mathcal{X}_k^{(0)}}(y-x)^2
\right),
\]
}
where $x$ are the symbol-bits in the corresponding subset.
%and $y$ is the received symbol.
%{\color{red} STEFANOS: what is y and what is y-x ?}

\textbf{Example: calculating bit-LLR for a received PAM4 symbol.} 
Assume that the receiver observes a signal of \m{y=\dc{-45.0}\,\mathrm{mV}}, corresponding to the first symbol ([01]=[-d]=[-50mv]) in the noise modeling example in~\autoref{sec:ber}.

\iffalse %% EXAMPLE MOVED TO APPENDIX

For the \textbf{MSB LLR}, the symbol subsets are bit-1: \m{\{+1d,+3d\}} and bit-0: \m{\{-3d,-1d\}}. With \m{d=50\,\mathrm{mV}} and \m{y=-45.0\,\mathrm{mV}} (so \m{y/d=-0.9}), we calculate the MSB's LLR that \m{y} represents as: 
\m{ 
\min_{x\in\{+1,+3\}}(y/d-x)^2=(-0.9-1)^2=\dc{3.61};
\min_{x\in\{-3,-1\}}(y/d-x)^2=(-0.9+1)^2=\dc{0.01}.
} 
This yields: 
\m{
(y/d-1)^2-(y/d+1)^2=\dc{3.60}. 
}
As discussed in~\autoref{sec:ber}, under AWGN, the normalized noise variance is
\m{
\sigma^2=\frac{E_s}{\mathrm{SNR}_{\mathrm{eff}}}
=\frac{5}{10^{\mathrm{SNR}_{\mathrm{eff,dB}}/10}}
\footnote{In \name{}, \(\mathrm{SNR}_{\mathrm{eff,dB}} \approx \dc{19.0}\), giving normalized \(\sigma^2 \approx \dc{0.064}\).},
}
so the MSB LLR via the max-log confidence is
\underline{
\m{
L_{\text{ch}}(\text{MSB})\approx \frac{3.60}{2\sigma^2}
\approx \dc{22.8}.
}
}

For the \textbf{LSB LLR}, the symbol subsets are bit-1: \m{\{-1d,+1d\}} and bit-0: \m{\{-3d,+3d\}}.
With \m{y/d=-0.9}, we calculate the LSB's LLR as
\m{
\min_{x\in\{-1,+1\}}(y/d-x)^2=(-0.9+1)^2=\dc{0.01},
\min_{x\in\{-3,+3\}}(y/d-x)^2=(-0.9+3)^2=\dc{4.41},
}
leading to
\m{
(y/d+1)^2-(y/d+3)^2 = \dc{-4.40}.
}
Under AWGN and \m{\sigma^2 \approx \dc{0.079}}, we have
\underline{
\m{
L_{\text{ch}}(\text{LSB})\approx \frac{-4.40}{2\times\sigma^2}\approx \dc{-27.8}.
}
}

\fi %% EXAMPLE MOVED TO APPENDIX

The received PAM4 symbol yields channel LLRs \m{L_{\text{ch}}=[\dc{+22.8},\,\dc{-27.8}]} for \m{\{\text{MSB},\,\text{LSB}\}}, which serve as soft-confidence inputs to the FEC decoder. 
% \dc{The complete calculation can be found in~\autoref{sec:app:llr}}. 
As detailed later in~\autoref{sec:fec:decode}, the decoder first produces a bit decision of \texttt{[01]}, which is then verified by the syndrome function using the same encoding matrix \m{H} introduced in~\autoref{sec:fec:encode}. This verification confirms that no error is present in this example, which is expected, as the injected noise of \m{+5.0\,\mathrm{mV}} is not sufficient to shift the signal into an adjacent symbol voltage level. Otherwise, the decoder initiates the error-correction process to repair the detected error.

%%%%%%%%%%%%%%%%%%%%%%%%%%%%%%%%%%%%%%%%

%\subsection{Layered min-sum QC-LDPC decoding}
\subsection{Decoding}
\label{sec:fec:decode}

\textbf{FEC decoding pipeline.}
%The receiver performs FEC decoding and error correction~\cite{ru2008}, reconstructs flits, and forwards them to the downstream router.
\name{} adopts a low-latency layered min-sum QC-LDPC decoder~\cite{fec_decode}, chosen for hardware efficiency and fast convergence.
Let the parity-check matrix \m{H} have \m{m} rows (layers), where each row corresponds to a check node \m{\mathrm{CN_i}} that enforces a group of parity constraints over a subset of variable nodes \m{\mathrm{v_j}}.

\textbf{Initialization.}
At the beginning of iteration \m{t}, each \m{\mathrm{v_j}} initializes its LLR from the channel input at the receiver:
\m{
L(v_j) \leftarrow L_\text{ch}(v_j).
}
The decoder then sweeps the layers sequentially (\m{i=0,\ldots,m{-}1}), performing check-node and variable-node updates in a layered schedule.

\textbf{Check-node update (exclude-self rule).}
For each edge \m{(CN_i,v_j)}, the outgoing message is:

{\footnotesize
\[
m_{cn_i \to v_j} \;=\;
\left(
\prod_{v_k \in \mathcal{N}(cn_i)\setminus v_j}
\mathrm{sgn}\!\bigl(m_{v \to cn_i}[v_k]\bigr)
\right)
\min_{v_k \in \mathcal{N}(cn_i)\setminus v_j}
\!\bigl|m_{v \to cn_i}[v_k]\bigr| \, .
\]
}

The sign is the product of all \emph{other} incoming \m{v\to c} signs, and the magnitude is the minimum of their absolute values.

\textbf{Variable-node update (layered accumulation).}
After computing \m{m_{cn_i \to v_j}}, \m{\mathrm{v_j}} updates its LLR incrementally:
\[
L(v_j) \leftarrow L(v_j) + m_{cn_i \to v_j},
\]
Subsequent layers immediately reuse the updated LLRs, accelerating convergence compared to flooding, where all layers use previous iteration's LLRs and new LLRs are updated at once after all $m_{cn \to v}$ for the current layer are calculated.

\textbf{Hard decision and syndrome check.}
After all layers are processed, the decoder decides each bit:
\[
\hat{c}_i =
\begin{cases}
0, & L(v_i)\ge 0, \\
1, & L(v_i)< 0,
\end{cases}
\qquad
\text{and verifies } H\hat{\mathbf{c}}^{T} \equiv \mathbf{0} \pmod{2}.
\]
If the syndrome is zero, decoding terminates successfully; otherwise, the decoder performs another layered sweep until either 1) the syndrome becomes zero---indicating all errors are corrected---or 2) a predefined iteration budget is reached, in which case the flit/packet is retransmitted.

% \begin{lstlisting}[language=C++]
%   /* Input Unit of Router */
%   std::vector<flit*> latch ; // Packet (Data + Parity or Control + Parity)
%   if (flit_t->get_type() == TAIL_PARITY_ || 
%       flit_t->get_type() == HEAD_TAIL_PARITY_){
%       bool corrputed = flase;
%       corrupted = SerDes->apply_decode();
%       if (corrupted){
%            SerDes->apply_repair();
%            /* 
%                 Extra cycles for repairing packet
%             */
%       }else{
%            put_in_input(); // Remove Parity and add to inputbuffer
%       }
%   }else{
%       SerDes->put_to_latch(); // Wait unitl parity tail flit.
%   }
% \end{lstlisting}

\textbf{Example: layered min-sum decoding.}
As introduced in~\autoref{sec:ber}, after transmission across the inter-chiplet channel, the receiver observes serial noisy PAM4 symbols
\m{
\mathbf{y}=\mathbf{x}+\mathbf{n}
=\underline{[\dc{-45.0},\,\dc{-171.0},\,\dc{+137.0},\,\dc{+158.0}]\,\mathrm{mV}}.
}
Each symbol is converted to bit-wise LLRs per~\autoref{sec:llr}, yielding
\m{
L_{ch} = [v_0, v_1, ..., v_7] = 
\bigl[
\underline{\dc{+22.8},\,\dc{-27.8},\,\dc{+122.5},\,\dc{+35.9},\,
\dc{-88.1},\,\dc{+18.7},\,\dc{-109.4},\,\dc{+29,4}}
\bigr].
}
For illustration, we consider \m{v_0\text{--}v_5} as data bits and \m{v_6,v_7} as parity bits\footnote{This simplified example captures the essential layered decoding schedule and applies directly to the 128-bit flit granularity in \name{}.}.
% We employ layered min-sum with \m{\alpha=0.75}~\cite{pcie6_fec_qa} and 
We use the following representative \m{3\times8} parity-check matrix (rows = CNs, columns = VNs):
\[
\small
H =
\begin{bmatrix}
      & v_0 & v_1 & v_2 & v_3 & v_4 & v_5 & v_6 & v_7\\
\mathrm{CN_0} & 1 & 1 & 0 & 1 & 0 & 0 & 1 & 0\\
\mathrm{CN_1} & 0 & 1 & 1 & 0 & 1 & 0 & 0 & 1\\
\mathrm{CN_2} & 1 & 0 & 1 & 0 & 0 & 1 & 0 & 0
\end{bmatrix}.
\label{eq:exampleH}
\]

In each layer, every CN updates its connected VNs (where there is a ``1'') using the current LLRs, which are incrementally refined within the same layer. Subsequent layers immediately reuse these updated values to accelerate convergence. 

% EXAMPLE MOVED TO APPENDIX
%%%%%%% Decoding %%%%%%%

%\textbf{Hard decision.}
After one full layered iteration (three layers), the updated variable-node LLRs are:
\[
L =
[\dc{69.3},\,\dc{-80.0},\,\dc{170.6},\,\dc{58.7},\,\dc{-117.5},\,\dc{69.3},\,\dc{-132.2},\,\dc{80.0}\,].
\]
The decoder then performs hard decisions according to:
\[
\hat{c}_i =
\begin{cases}
0, & L(v_i) \ge 0,\\
1, & L(v_i) < 0,
\end{cases}
\quad \Rightarrow \quad
\hat{\mathbf{c}} = [0,1,0,0,1,0,1,0].
\]
Finally, the syndrome check \(H\hat{\mathbf{c}}^{T}=0\pmod{2}\) confirms that all parity constraints are satisfied. In this example, the codeword passes the check (we receive the same message as before, encoded as described in \autoref{sec:ber}), and decoding successfully converges after a single layered iteration. If the syndrome is non-zero, the decoder reuses the updated LLRs and repeats until either the syndrome becomes zero or a limit is reached and a packet is retransmitted.

\begin{figure}[!t]
  \centering
  \includegraphics[width=0.5\textwidth]{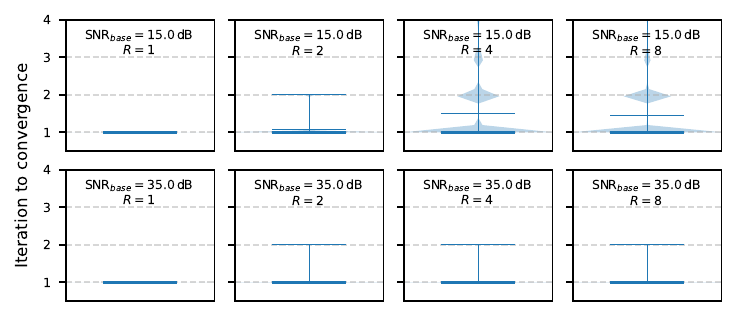}
  \caption{Sensitivity of QC-LDPC convergence to varying base SNR and parity-byte configurations.
  \review{E.Other}
  \rebut{Takeaway: At the baseline 35\,dB SNR, FEC decoding converges within 2 iterations for all parity-byte settings $R$.}
  }
  \label{fig:violin}
  \tv{}
\end{figure}

\textbf{FEC-decoder latency.}
We synthesize a layered min-sum QC-LDPC decoder in Verilog. Because the decoding process is runtime-determined, we set the decoder loop budget empirically: \autoref{fig:violin} reports violin plots of \emph{iterations to convergence}\footnote{We exclude error-free flits and runs that do not converge within a generous cap of 20 iterations.} versus parity bytes $R=1,2,4,8$ across link qualities \m{\mathrm{SNR}_{\text{base}}=\dc{15.0}\,\mathrm{dB}} and \m{\dc{35.0}\,\mathrm{dB}}. 
At the baseline \m{\mathrm{SNR}_{\text{base}}=\dc{35.0}\,\mathrm{dB}}, all tested code rates converge within \m{\le 2} iterations. 
We thus cap the decoding loop budget at \m{N=4}---bounding worst-case latency while retaining margin for lower \m{\mathrm{SNR}_{\text{base}}}. For both CCD and the IOD, the syndrome stage uses \m{L_{\text{syndrome}}=1} cycle, and each decoding iteration uses \m{L_{\text{iter}}=1} cycle (including all layers of LLR updates and comparisons). The total latency after \m{N} iterations is:
\[
\text{Latency}(N) = (N+1)L_{\text{syndrome}} + N L_{\text{iter}} = 2N+1\ \text{cycles}, N \le 4.
\]

At a noisier channel of \m{15.0\,\mathrm{dB}}, we observe 2 phenomena. First, with fewer parity bytes (\eg{}, $R=1$ or $R=2$), many errors cannot be corrected by FEC because the parity-byte budget is insufficient; among those that \textit{are} detected, 2 iterations are sufficient for correction. 
Second, increasing the number of parity bytes enables correction of more errors, but requires additional decoding iterations to convergence.

% \emph{Note:} If a larger figure (e.g., 61 cycles) is reported elsewhere, it implies additional per-iteration pipeline stages or layer-unrolled passes beyond \(L_{\text{CN}}=2\); with the parameters above, the closed-form \(3N+1\) applies.

%%%%%%%%%%%%%%%%%%%%%%%%%%%%%%%%%%%%%%%%

\subsection{Putting it all together: \phy{} router microarchitecture}
\label{sec:router-micro}

\begin{figure}[!t]
  \begin{center}
    \includegraphics[width=0.5\textwidth]{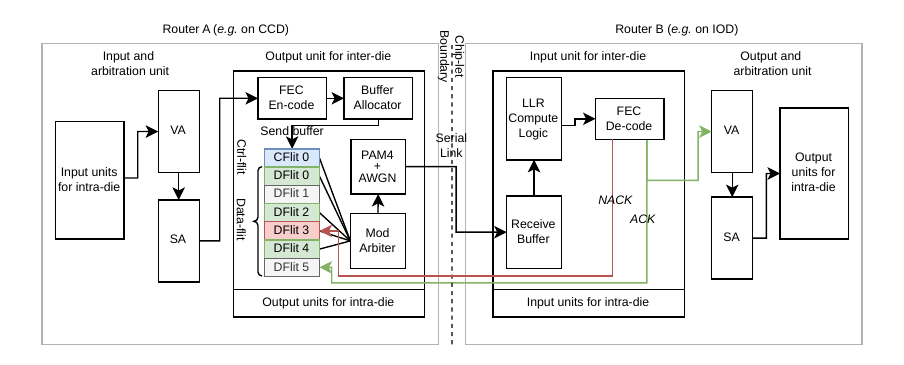}
  \end{center}
  \caption{Microarchitecture for sender (Router A) and receiver (Router B) \phy{} routers.}
  \label{fig:router-a-b}
  \tv{}
\end{figure}

To support the mechanisms described from~\autoref{sec:fec:encode} to \autoref{sec:fec:decode}, \name{} integrates \phy{}-link-level flow control into the router microarchitecture at chiplet boundaries. We illustrate the mechanism with~\autoref{fig:router-a-b}, where Router~A (sender) and Router~B (receiver) reside on separate chiplets and communicate over a serialized inter-chiplet link. We next detail the end-to-end flow control process, from encoding at the sender, to inter-chip link traversal, then to decoding at the receiver.

% \begin{lstlisting}[language=C++]
%   /* Output Unit of Router */
%   std::vector<flit*> latch ; // Packet (Data or Control Packet)
%   if (flit_t->get_type() == TAIL_ || flit_t->get_type() == HEAD_TAIL_){
%       SerDes->apply_encode(latch);
%   }else{
%       SerDes->put_to_latch(flit_t); // Wait unitl tail flit.
%   }
% \end{lstlisting}

\textbf{1, Encoding.}
As illustrated in~\autoref{fig:router-a-b}, on the sender side (Router~A), \name{} implements an inter-die output unit, which integrates: 1) an FEC encoder, 2) a send buffer and buffer allocator, 3) a modulation arbiter, and 4) a modulator. As introduced in~\autoref{sec:fec:encode}, \name{} performs encoding/decoding at \emph{flit}-level granularity to minimize hardware overhead: for data packets, each of the 6$\times$128-bit flits (1~\textsf{HEAD}, 4~\textsf{BODY}, 1~\textsf{TAIL}) is encoded independently, whereas a control packet consists of a single encoded \textsf{HEAD-TAIL} flit.

\textbf{2, Inter-chiplet flow control and link traversal.}
After FEC encoding (\autoref{sec:fec:encode}), each flit (data + parity) is queued in the send buffer and arbitrated for modulation/serialization on the cross-chiplet link (\autoref{sec:llr}). To balance allocation flexibility with implementation simplicity, \name{} adopts \textit{cut-through} flow control. 
1) The send buffer comprises flit-sized entries that are reserved at \textit{packet} granularity (1 entry for control packets, 6 entries for data packets). 
2) After link traversal, the receiver returns per-flit ACK/NACKs to the sender. This flit-level credit feedback localizes recovery, as only flits that fail decoding are resent by the sender. 
3) A packet's reservation in the send buffer is released only after \emph{all} of its flits have been ACKed. 
4) Multiple packets can coexist in the send buffer; if one packet stalls, others can still advance through modulation arbitration, sustaining high link utilization and avoiding head-of-line blocking.

\textbf {3, Decoding.} After link traversal through the noisy channel (\autoref{sec:ber}), Router~B accumulates soft samples (data+parity) into a receive buffer, then invokes LLR compute logic and FEC decoder with a bounded iteration budget (\autoref{sec:fec:decode}). On a successful decode, the flit is delivered for onward transmission and an ACK is returned to Router~A. On failure, Router~B issues a NACK, which triggers re-modulation and re-transmission of that affected flit.

\section{Evaluation}

\subsection{Methodology}
\label{sec:metho}

\begin{table}[!t]
  \centering
  \footnotesize
  \caption{gem5 configurations}
  \begin{tabular}{ p{3.2cm}|p{4.4cm}  }
    \Xhline{3\arrayrulewidth}
    \review{C.Other}
    \textit{\textbf{
    \rebut{Microarchitecture}}
    } & \textit{\textbf{}} \\
    \Xhline{3\arrayrulewidth}
    Core/Reorder buffer (ROB) & 3.0 GHz x86\_64 OoO, \dc{512} entries\\
    \rebut{Instruction queue (IQ)} & \dc{160} entries\\
    \rebut{Load/store-queue (LQ/SQ)} & \dc{192}, \dc{114} entries\\
    Int/Float/Vec Register & \dc{332}/\dc{332}/\dc{332} entries \\
    Fetch/Decode/Rename & 8 micro-ops per cycle \\
    Issue/Squash/Commit & 8 micro-ops per cycle \\
    Branch predictor/BTB & 64 Kbits TAGE\_SC\_L/4096 entries \\
    MemDep predictor & 1024 entries LFST and SSIT store set \\
    & History cleaned every 250,000 cycles \\ 
    \Xhline{3\arrayrulewidth}
    \textit{\textbf{\rebut{Cache configuration}}} & \textit{\textbf{}} \\
    \Xhline{3\arrayrulewidth}
    L1 I/DCache & 64KB-I/64KB-D, 8 ways, 64\,Byte cacheline \\
    L2 Cache (LLC) & 1MB/bank, 16 ways, 64\,Byte cacheline \\
    Cache coherence & MESI-Two-Level \\
    \Xhline{3\arrayrulewidth}
    \textit{\textbf{Chiplet configuration}} & \textit{\textbf{}} \\
    
    \Xhline{3\arrayrulewidth}
    CCD layout & \dc{8} cores/CCD (\dc{2$\times$4}), \dc{4} CCDs, \dc{32} cores \\
    
    CCD NoC & \dc{1}-cycle \dc{2.0}\,GHz router, \dc{2}-cycle \dc{2.0}\,GHz link, \dc{128}-bit link bandwidth, \dc{1} \phy{}-router per CCD \\
    
    IOD layout & \dc{4} \phy{}-routers (\dc{2$\times$2}), \dc{8} memory ctrls \\
    
    IOD NoC & \dc{1}-cycle \dc{1.0}\,GHz router, \dc{2}-cycle \dc{1.0}\,GHz link, \dc{128}-bit link bandwidth, \dc{2} mem ctrls per router \\
    
    Channel noise/Modulation & \dc{AWGN} channel, \dc{PAM4} modulation \\
    
    FEC encoding & Code ratio R=\dc{0.88}, Z=\dc{8}, \dc{16} parity bits, 1-cycle encoding latency, 1-cycle modulation latency \\
    
    FEC decoding & Layered Min-Sum, \dc{1} cycle syndrome calculation, \dc{4}-iteration budget, \dc{1} cycle per iteration \\
    
    Symbol rate and SNR & 
    symbol rate: 32\,GT/s. Jitter: \dc{26.0}\,dB, 
    crosstalk: \dc{20.0}\,dB, \m{\mathrm{SNR}_{\mathrm{base}}}: \dc{35.0}\,dB\\
    % Error distribution & AWGN Normal distribution\\
    \Xhline{3\arrayrulewidth}
    \textit{\textbf{Mem. configuration}} & \textit{\textbf{}} \\
    \Xhline{3\arrayrulewidth}   
    Memory & 32GB DDR5, 4400 GHz, 8 controllers\\
    \Xhline{3\arrayrulewidth}
  \end{tabular}
  \label{tab:simulations}
\end{table}

\textbf{Simulation.} 
We implement \name{} by integrating all \phy{} components—FEC encoding/decoding, channel noise modeling, modulation/demodulation, LLR computation, \etc{}—into gem5 Garnet~\cite{garnet}. We evaluate \name{} in gem5 system emulation (SE) using x86\_64 out-of-order cores that implement the architecture shown in~\autoref{fig:intro}. The core, uncore, and CCD/IOD parameters are summarized in~\autoref{tab:simulations}.

\begin{table}[t]
\centering
\footnotesize
\caption{
\review{B.Other, C.Q2, C.Other, D.Q1}
\rebut{Parameter settings in \name{}}
}
\label{tab:modeling_validation}
{\color{rebut}
\begin{tabular}{p{1.1cm} p{0.9cm} p{5.1cm}}
\toprule
\textbf{Parameter} & \textbf{Value} & \textbf{Reference/Source} \\

\midrule
Parity
& 2 Bytes per 16-Byte flit
& Empirically evaluated in~\autoref{fig:waterfall_all}. \name{} is compatible with UCIe formats such as the 68B format, where its FEC bytes can be injected into the unused bytes~\cite{ucie_spec}. \\

\midrule
Symbol rate
& 4\,GT/s - 32\,GT/s
& $1\,\mathrm{GT/s}$ denotes $10^{9}$ symbol transmissions per second. 
The maximum symbol rate supported by UCIe~2.0 is up to $32\,\mathrm{GT/s}$ per SerDes lane~\cite{ucie_spec}. \\

\midrule
SNR$_{\mathrm{base}}$
& $\approx$ 35~dB 
& IEEE HIR 2024 (Chapter~2, HPC) reports that $\sim$35~dB channel quality is typical for very-short-range inter-chiplet SerDes IO~\cite{IEEE_HIR_HPC_2021}. \\

\midrule
Jitter
& $\approx$ 1\,ps
& PCI-SIG indicates that high-speed reference clock and differential jitter are on the order of hundreds of femtoseconds (\eg{}, 0.7 ps for PCIe 5.0). \name{} uses 1\,ps to tolerate more electrical budgets~\cite{pcie5_qa}. \\

\midrule
Crosstalk
& $\approx$ 20~dB 
& UCIe standard’s guidance for signal integrity (SI) indicates that crosstalk for 32~GT/s lanes is approximately $20$~dB~\cite{UCIeSignalIntegrity}. \\

\bottomrule
\end{tabular}
\tv{}
} % End of blue color

\end{table}

{\color{rebut}
\textbf{Fidelity of \phy{}-link modeling in \name{}.}
\review{C.Q2, D.Q1}
Unlike prior simulators that assume fixed link latencies, 
%and approximate channel noise using simple bit-flips, 
\name{} explicitly models PAM4 modulation,
injects AWGN-based noise to signal 
symbols~\cite{9401125}, and performs soft-decision FEC decoding based on log-likelihood ratios~\cite{QC_LDPC,fec_decode}. This approach is reflected in the IEEE Heterogeneous Integration Roadmap (HIR) 2024~\cite{IEEE_HIR_HPC_2021}, which identifies increased channel noise, the adoption of PAM4 to sustain high bandwidth, tighter signal crosstalk and jitter margins, and growing reliance on FEC, as first-order challenges in emerging chiplet systems~\cite{synopsis}. 
We take two steps to ensure model fidelity of \name{}. First, all parameters in \name{} (\eg{}, channel SNR, signaling rates, crosstalk, and jitter) are aligned with publicly available specifications and industry datasheets, as summarized in~\autoref{tab:modeling_validation}. 
Second, we validate \name{} against three chiplet-based commercial processors, with validation results presented in~\autoref{sec:vali}.

%{\color{red} REMOVE: 
%In summary, \name{}, as outlined in~\autoref{fig:overview}, captures how \phy{}-link impairments affect error detection and correction, potentially triggering flit retransmissions and ultimately impacting system-level performance. These effects are simulated dynamically (with channel noise injection)---an aspect that fixed-latency models such as \textsf{HeteroGarnet} cannot capture.
%}
% As signaling rates scale toward 32--64\,GT/s~\cite{ucie_spec,ucie_spec_3}, inter-chiplet \phy{} links operate under increasingly constrained SNR budgets, making \phy{} reliability a system-level concern rather than a secondary implementation detail. 
}

\textbf{Benchmarks.}
We evaluate a diverse set of \dc{14} benchmarks spanning multiple suites, including \dc{3} programs from GAPBS~\cite{gapbs}, \dc{3} from SPEC CPU2017~\cite{spec17}, \dc{5} from Splash~4~\cite{splash4}, \dc{2} from Rodinia~\cite{rodinia}, and \dc{the} XSBench~\cite{xsbench}.
\autoref{fig:char_programs} summarizes the program characteristics in terms of instructions per cycle (IPC), which indicates whether a workload is memory- or compute-bound, and last-level cache (LLC) misses per kilo-instruction (MPKI), which reflects the intensity of inter-chiplet communication.
Since an LLC miss triggers a CCD-to-IOD access, we classify applications based on LLC MPKI: values above 10 indicate inter-chiplet communication–dominated programs, between 1 and 10 represent moderate communication intensity, and below 1 correspond to compute-bound workloads.
% This classification is consistent with IPC trends---programs with higher MPKI exhibit lower IPC due to stalls caused by remote memory accesses, while compute-bound workloads achieve higher IPC with limited off-chip traffic.
In our evaluation, we spawn four instances of the same program and run one process on each of the four CCDs to mimic a typical multi-programming server environment. We create checkpoints after initialization phase (\ie{}, upon entering the ROI).

\begin{table}[!t]
\centering
\footnotesize
\caption{Benchmark characteristics.}
\begin{tabular}{llllc}
\toprule
\textbf{Suite} & \textbf{Program} & \textbf{Character} & \textbf{IPC} & \textbf{LLC MPKI} \\
\midrule
\multirow{3}{*}{GAPBS~\cite{gapbs}}
 & bc         & inter-chiplet  & \cellblue{30} 0.55 & \cellblue{25} 22.8  \\
 & bfs        & inter-chiplet  & \cellblue{5} 0.23  & \cellblue{75} 77.66 \\
 & cc         & inter-chiplet  & \cellblue{15} 0.42 & \cellblue{35} 35.48 \\
\midrule
\multirow{3}{*}{SPEC 2017~\cite{spec17}}
 & leela      & intra          & \cellblue{40} 1.05 & \cellblue{0} 0.09   \\
 & mcf        & inter-chiplet  & \cellblue{25} 0.58 & \cellblue{70} \dc{70.65} \\
 & omnetpp    & mixed          & \cellblue{35} 0.88 & \cellblue{5} 2.19   \\
\midrule
\multirow{5}{*}{Splash 4~\cite{splash4}}
 & lu-cb      & mixed         & \cellblue{40} 1.03 & \cellblue{5} 2.12   \\
 & ocean-cp   & inter-chiplet & \cellblue{20} 0.57 & \cellblue{20} 15.11 \\
 & radix      & mixed         & \cellblue{45} 1.27 & \cellblue{5} 2.68   \\
 & radiosity  & intra         & \cellblue{55} 1.79 & \cellblue{0} 0.56   \\
 & volrend    & mixed         & \cellblue{40} 1.20 & \cellblue{5} 1.86   \\
\midrule
\multirow{2}{*}{Rodinia~\cite{rodinia}}
 & kmeans     & mixed         & \cellblue{50} 1.50 & \cellblue{10} 5.02  \\
 & sc         & mixed         & \cellblue{50} 1.52 & \cellblue{5} 3.71   \\
\midrule
XSBench~\cite{xsbench} & XSBench & inter-chiplet    & \cellblue{5} \dc{0.25} & \cellblue{100} 122.8 \\
\bottomrule
\end{tabular}
\tv{}
\label{fig:char_programs}
\end{table}

\subsection{Evaluation Results}

%%%%%%%%%%%%%%%%%%%%%%%%%%%%%%%%%%%%%%%%%%%%%%%%

\subsubsection{Validation of \name{}}
\label{sec:vali}

To validate \name{}, we compare core-to-core (C2C) communication latency across \name{}, \textsf{HeteroGarnet} (denoted \hg{}), and an AMD EPYC~9454P (Zen~4) processor. We run \name{} and \hg{} in gem5 full-system mode to run Linux and the C2C benchmark~\cite{c2c}, which allows us to control CPU affinity and record C2C values using Linux kernel timestamps. For a fair comparison, we configure both \name{} and \hg{} to mirror the 9454P architecture, which consists of 8 CCDs, each integrating 6 out-of-order (OoO) cores, for a total of 48 cores. Due to space constraints, without loss of generality, we report C2C latency from all cores within CCD~0 to every other core in the system.

\begin{figure}[!t]
    \centering
    \includegraphics[width=0.5\textwidth]{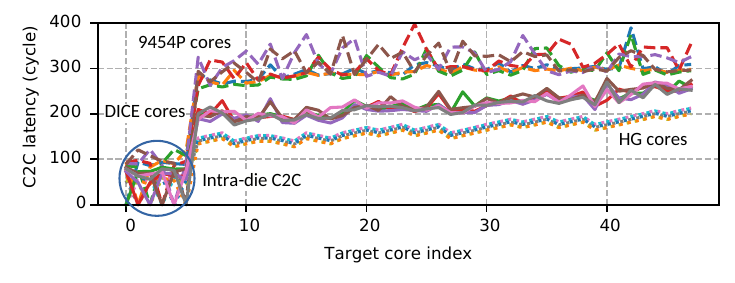}
    \caption{Validation of \name{} against \textsf{AMD EPYC 9454P} (8 CCDs $\times$ 6 cores):
    \review{F.Q4}
    \rebut{
    C2C max latency for all 6 cores within one representative CCD, measuring communication from each of these cores to every other of the 47 cores in the system. As shown, by modeling \phy{} dynamics, \name{} (with default parameters) narrows the gap and more faithfully emulates the latency variability of the actual processor than \hg{}.
    }
    }
    \label{fig:c2c}
    % \tv{}
\end{figure}

As shown in \autoref{fig:c2c}, we report the \textit{maximum} C2C latency for each source--destination core pair, as long-tail communication latency has a greater impact on application performance than average latency~\cite{sharifi2012addressing, farrokhbakht2021pitstop} (see also \autoref{sec:relevance}). Compared to \hg{}, \name{} more closely matches the C2C latency profile of the actual EPYC processor. %This improvement stems from \name{}'s more accurate modeling of inter-chiplet communication overhead, yielding a more faithful representation of latency in an actual chiplet-based system.
%{\color{purple}
\rebut{To quantitatively evaluate the difference, we report the Root Mean Square Error (RMSE), measured in cycles, from \hg{} and \name{} to the AMD EPYC 9454P processor. Excluding intra-die measurements, \hg{} gives an RMSE of \dc{141.2} cycles (\dc{46.4\%} of the average maximum latency, \dc{304.6} cycles, on the actual processor),\footnote{Whether we normalize by the average (mean) or the RMS of the reference curve makes very little difference to our results.} whereas \name{} achieves \dc{89.5} cycles (\dc{29.4\%}). \name{} therefore narrows the gap to actual hardware, improving the fidelity of the modeled tail latency by \dc{17.0\%} over \hg{}.}

\rebut{
\review{B.Q2}
\textbf{Additional validation with publicly available \emph{average} C2C latencies.} While it is the tail latency that is important for end performance (see \autoref{sec:relevance}), we perform additional validation using publicly available data~\cite{c2c} for the \emph{average} C2C latency.\footnote{Using publicly available C2C datasets prevents us from exploring maximum latency and restricts us to average latency.} We evaluate two additional architectures: AMD ThreadRipper 3960X (8 CCDs $\times$ 3 cores) and AMD EPYC 7R13 (6 CCDs $\times$ 8 cores). As summarized in \autoref{tab:rmse}, on the ThreadRipper 3960X, \name{} reduces RMSE from \dc{36.6} cycles (\dc{19.1\%}) with \hg{} to \dc{17.1} cycles (\dc{8.9\%}). On the EPYC 7R13, the error drops from \dc{39.9} cycles (\dc{18.9\%}) to \dc{24.9} cycles (\dc{11.8\%}), and on the EPYC 9454P from \dc{100.4} cycles (\dc{40.5\%}) to \dc{73.9} cycles (\dc{29.8\%}). Overall, \name{}, with default parameters,  consistently improves modeling fidelity, reducing relative RMSE by \dc{7.1\%}–\dc{10.7\%} compared to \hg{} across architectures.}

\rebut{
\review{B.Q1, C.Other}
Moreover, with enough samples, \name{} can be calibrated to fit a range of architectures/workloads, but the main limitation is the lack of reliable data from actual implementations, which currently prevents validation of the calibrated values at the level of individual parameters. 
As more chiplet-based systems are introduced and characterized, especially those adopting common standards such as UCIe~\cite{ISSCC2026}, such calibration should become 
more accurate to real hardware.
}

% \begin{table}[!h]
%   \centering
%   \color{purple}
%   \scriptsize
%   \caption{
%   \color{purple}
%   \review{\textbf{[NEW!]}}
%   RMSE (cycles) and Relative RMSE (in parentheses), normalized to the mean average C2C latency across all core pairs, for \name{} and \hg{} on three AMD chiplet-based processors.
%   }
%   \begin{tabular}{lcccc}
%     \toprule
%     \textbf{} & \textbf{ThreadRipper 3960X } & \textbf{EPYC 7R13} & \textbf{EPYC 9454P } \\
%     \textbf{} & \textbf{8 CCDs $\times$ 3 cores } & \textbf{6 CCDs $\times$ 8 cores} & \textbf{8 CCDs $\times$ 6 cores } \\
    
%     \midrule
%     \hg{}   & 38.3 cycle (20.2\%) & 34.4 cycle (20.0\%) & 100.4 cycle (39.7\%) \\
%     \name{} & 24.8 cycle (13.1\%) & 16.0 cycle (9.3\%) & 73.9 cycle (29.2\%) \\
%     \bottomrule
%   \end{tabular}
%   \label{tab:rmse}
% \end{table}

\begin{table}[t]
\centering
\scriptsize
\caption{\rebut{RMSE comparison}}
\color{rebut}
\review{B.Q2}
\setlength{\tabcolsep}{5pt}
\begin{tabular}{lcc cc cc}
\toprule
 & \multicolumn{2}{c}{ThreadRipper 3960X} 
 & \multicolumn{2}{c}{EPYC 7R13} 
 & \multicolumn{2}{c}{EPYC 9454P} \\

 & \multicolumn{2}{c}{8 CCDs $\times$ 3 cores} 
 & \multicolumn{2}{c}{6 CCDs $\times$ 8 cores} 
 & \multicolumn{2}{c}{8 CCDs $\times$ 6 cores} \\

\cmidrule(lr){2-3} \cmidrule(lr){4-5} \cmidrule(lr){6-7}
 & Avg & RMSE & Avg & RMSE & Avg & RMSE \\
\midrule

\hg{}   & 158.8 & 36.6 (19.1\%) & 180.5 & 39.9 (18.9\%) & 152.6 & 100.4 (40.5\%) \\
\name{} & 186.8 & 17.1 (8.9\%) & 205.4 & 24.9 (11.8\%)  & 177.8 & 73.9 (29.8\%) \\

\bottomrule
\end{tabular}
\label{tab:rmse}
\tv{}
\end{table}

\begin{figure}[!t]
  \centering
  \includegraphics[width=0.475\textwidth]{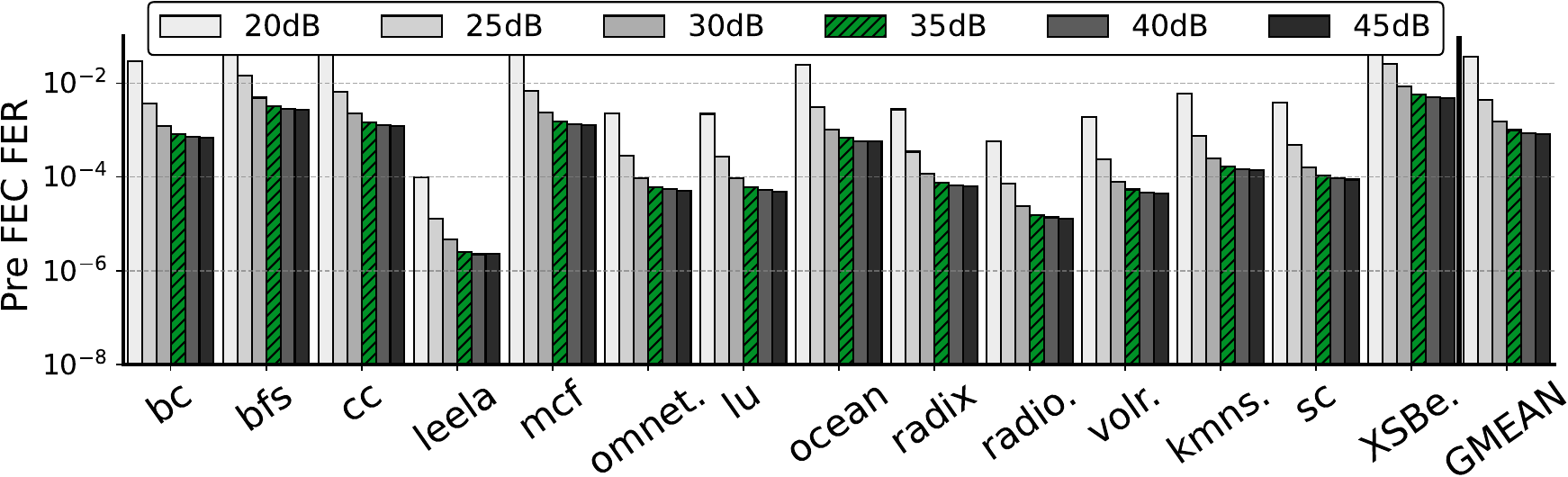}
  \caption{Pre-FEC FER based on varied \m{\mathrm{SNR}_{\mathrm{base}}}.}
  \label{fig:pre-fec:lab}
  % \tv{}
\end{figure}

\begin{figure}[!t]
  \centering
  \includegraphics[width=0.475\textwidth]{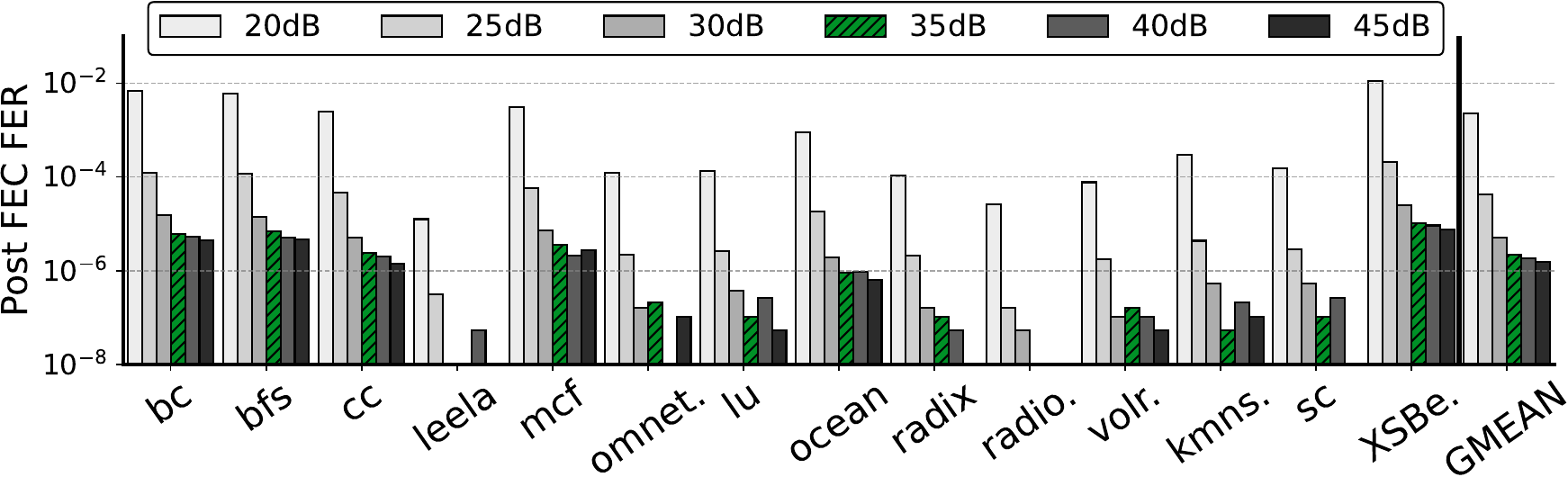}
  \caption{Post-FEC FER based on varied \m{\mathrm{SNR}_{\mathrm{base}}}.}
  \label{fig:post-fec:lab}
  \tv{}
\end{figure}

% \begin{figure}[!t]
%   \centering
%   \includegraphics[width=0.48\textwidth]{figs/FinalResults/Base_SNR_norm-crop.pdf}
%   \caption{\red{Number of Post-FEC errors (including false-positive) based on varied \m{\mathrm{SNR}_{\mathrm{base}}}.}}
%   \label{fig:post-fec:modu}
% \end{figure}

\subsubsection{Impact of \m{\mathrm{SNR}_{\mathrm{base}}} on Pre- and Post-FEC FER}
\label{sec:eva-1}

As discussed in~\autoref{sec:awgn}, flit-error ratio (FER) is governed by the effective SNR (\m{\mathrm{SNR}_{\mathrm{eff}}}), which combines the baseline SNR (\m{\mathrm{SNR}_{\mathrm{base}}}) with jitter and crosstalk via~\autoref{eq:snr_eff}. While jitter and crosstalk are primarily determined by channel characteristics (\eg{}, link frequency and distance between wires), \m{\mathrm{SNR}_{\mathrm{base}}} drifts with runtime operating conditions (\eg{}, thermal conditions). We next analyze \m{\mathrm{SNR}_{\mathrm{base}}} to quantify its impact on data transmission reliability, and report the pre-FEC and post-FEC FER in~\autoref{fig:pre-fec:lab} and~\autoref{fig:post-fec:lab}.

From \autoref{fig:pre-fec:lab} we can make 3 observations. 1) Communication-intensive applications such as \textsf{bfs}, \textsf{cc}, and \textsf{XSBench} suffer more from errors, as inter-chiplet communication occurs more frequently. 2) For each application, FER varies significantly with \m{\mathrm{SNR}_{\mathrm{base}}}. At low \m{\mathrm{SNR}_{\mathrm{base}}} values (\dc{20--25\,dB}), all workloads experience substantial pre-FEC errors. As \m{\mathrm{SNR}_{\mathrm{base}}} increases, errors drop sharply, revealing a non-linear relationship between channel quality and FER. 3) Beyond \dc{35\,dB}, further increases in \m{\mathrm{SNR}_{\mathrm{base}}} yield only marginal improvement, suggesting \dc{35\,dB} as a practical operating point.

\autoref{fig:post-fec:lab} shows the post-FEC FER for different applications. In \name{}, using FEC instead of relying solely on conventional error detection via CRC (\eg{}, in AMD's Infinity Fabric~\cite{beck2018zeppelin}, Intel CXL~\cite{CXL_3_1}, and a recent work of RXL~\cite{scaling_out}) allows most errors to be corrected online, leaving only post-FEC errors subject to retransmission. As shown in the figure, compared to~\autoref{fig:pre-fec:lab}, \name{} corrects on average \dc{97.8\%} of the errors, leaving only \dc{2.2\%} of the original errors for retransmission. This significantly reduces both runtime overhead and energy consumption, as flits are corrected in place rather than being resent. 

\begin{figure}[t]
  \begin{center}
    \includegraphics[width=0.47\textwidth]{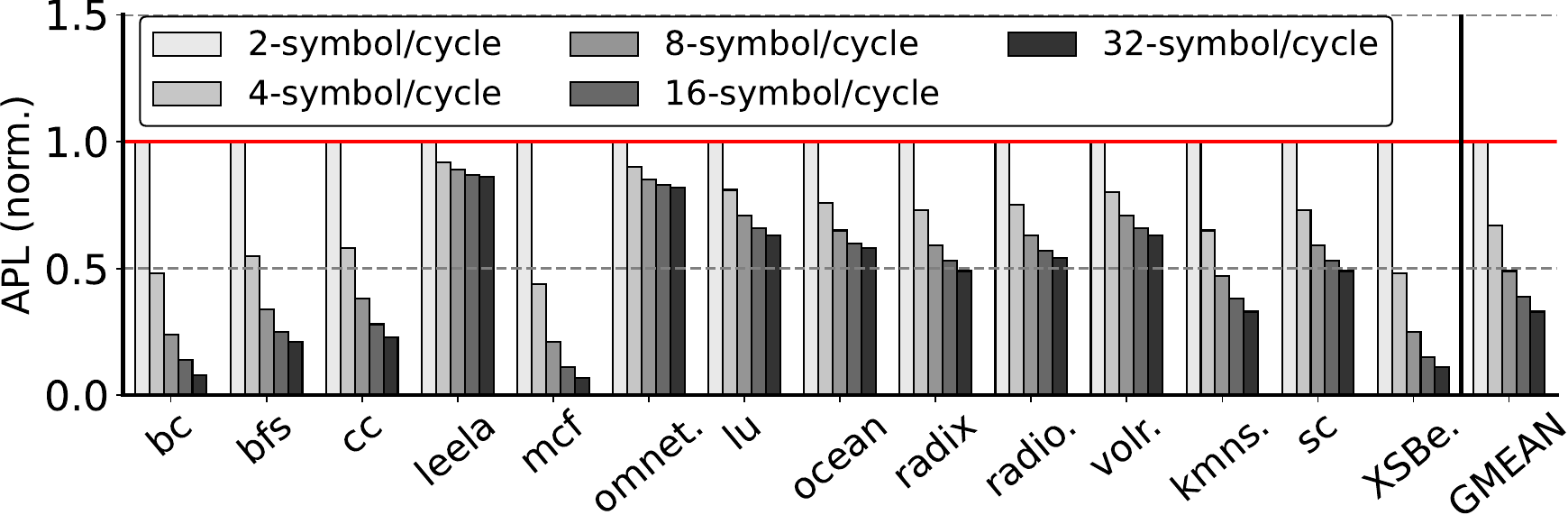}
  \end{center}
  \caption{Normalized avg packet latency across symbol rates.}
  \label{fig:apl_icl}
  % \tv{}
\end{figure}

\begin{figure}[t]
  \begin{center}
    \includegraphics[width=0.47\textwidth]{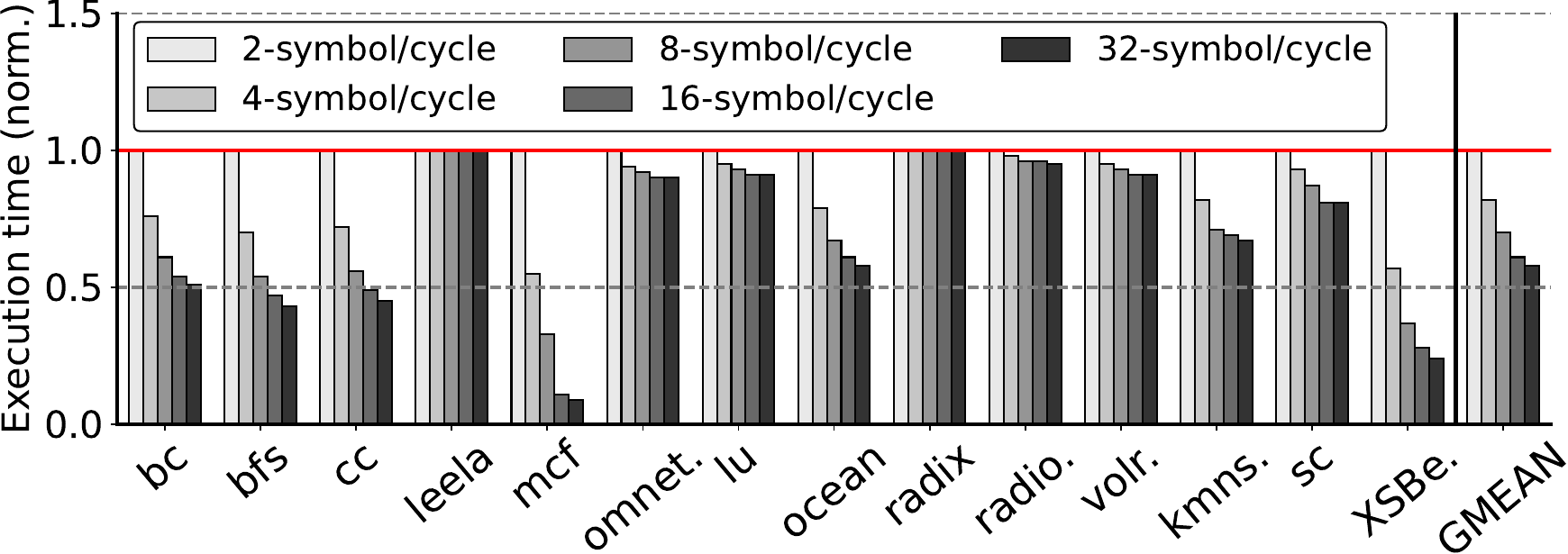}
  \end{center}
  \caption{Normalized execution time across symbol rates.}
  \label{fig:exe_icl}
  \tv{}
\end{figure}

\subsubsection{Average packet latency and application runtime across inter-chiplet link bandwidth} 

We evaluate the impact of varying \phy{} link symbol rates (2–32 symbols/cycle) on overall application average packet latency (APL) and execution time (both are normalized to 2-symbol/cycle). As shown in~\autoref{fig:apl_icl}, as expected, increasing the link symbol rate substantially reduces average packet latency, with diminishing returns however beyond 16 symbols/cycle as serialization delay and queuing effects become less dominant. This latency improvement translates directly to higher application performance: in~\autoref{fig:exe_icl}, total execution time decreases consistently with higher symbol rates, particularly for communication-intensive workloads (\eg{} \textsf{bc}, \textsf{bfs}, \textsf{cc}, \textsf{mcf}, and \textsf{XSBench}). Compute-bound programs (\eg{}, \textsf{leela}, \textsf{radiosity}) show marginal improvement, indicating limited sensitivity to inter-chiplet link bandwidth.

% 4 %%%%%%%%%%%%%%%%%%%%%%%%%%%%%%%%%%%%%%%%%%%%%%%

\subsubsection{Average packet latency and application runtime across IOD router latency}

\begin{figure}[t]
  \begin{center}
    \includegraphics[width=0.44\textwidth]{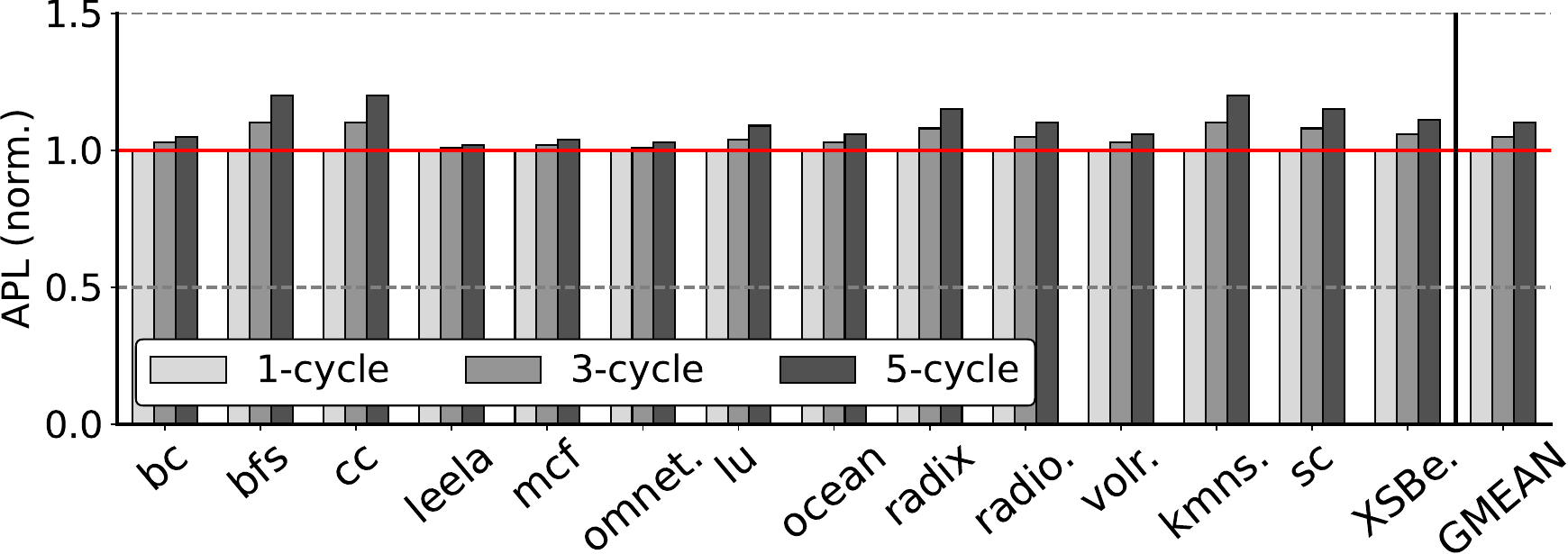}
  \end{center}
  \caption{Norm. avg packet latency across IOD link latency.}
  \label{fig:APL_IOD}
  % \tv{}
\end{figure}

\begin{figure}[t]
  \begin{center}
    \includegraphics[width=0.47\textwidth]{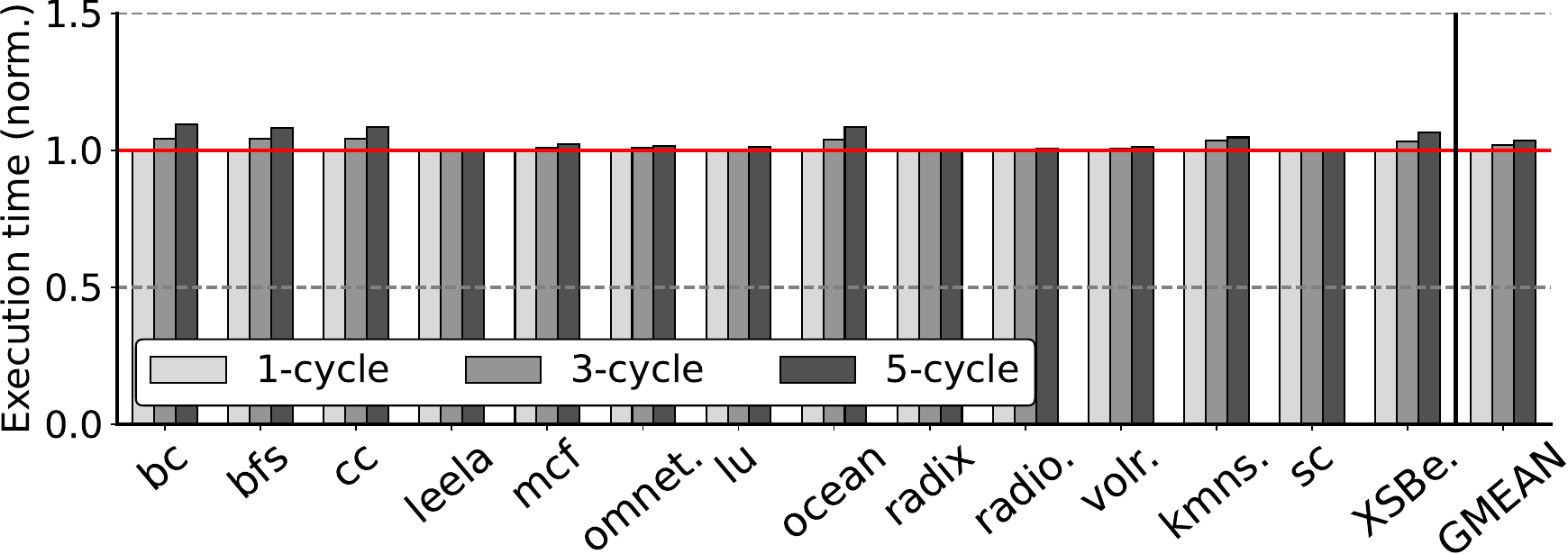}
  \end{center}
  \caption{Normalized execution time across IOD link latency.}
  \label{fig:EXE_IOD}
  \tv{}
\end{figure}

A fundamental difference between CCD and IOD is that IOD is often manufactured in a less advanced technology node (\eg{}, 14 nm vs. 5 nm for the CCDs in the AMD EPYC 7002 series~\cite{AMD_EPYC}). We next examine the effects of the IOD having different router latencies to mimic this technology gap, as a slower IOD router can backlog packets that travel cross chiplet boundaries and downgrade application performance.
\autoref{fig:APL_IOD} and~\autoref{fig:EXE_IOD} report normalized average packet latency and application execution time with varying the IOD router latency. We observe that, for compute-bound applications such as \textit{leela} and \textit{radio}, the IOD router latency has only a minor effect on both average packet latency and application runtime. This is because the amount of traffic these applications send through the IOD is relatively small compared to communication-intense applications. As a result, increasing the IOD router latency only delays a small fraction of packets and thus has limited impact on overall performance. In contrast, for benchmarks with more intensive inter-chiplet communication, such as \textsf{bfs}, \textsf{cc}, IOD router latency variation results in more pronounced performance degradation.

%%%%%%%%%%%%%%%%%%%%%%%%%%%%%%%%%%%%%%%%%%%%%%%%%

% 3 %%%%%%%%%%%%%%%%%%%%%%%%%%%%%%%%%%%%%%%%%%%%%%%

% \begin{figure}[t]
%   \centering
%   \begin{subfigure}{.47\linewidth}
%     \centering
%     \includegraphics[width=\linewidth]{figs/text/yuan/cc0.pdf}
%     \caption{\scriptsize Cross-chiplet flow: \textbf{2} ctrl, \textbf{2} data.}
%     \label{fig:cc:0}
%   \end{subfigure}
%   \begin{subfigure}{.47\linewidth}
%     \centering
%     \includegraphics[width=\linewidth]{figs/text/yuan/cc1.pdf}
%     \caption{\scriptsize Cross-chiplet flow: \textbf{2} ctrl, \textbf{2} data.}
%     \label{fig:cc:1}
%   \end{subfigure} \\
%   \begin{subfigure}{.47\linewidth}
%     \centering
%     \includegraphics[width=\linewidth]{figs/text/yuan/cc2.pdf}
%     \caption{\scriptsize Cross-chiplet flow: \textbf{4} ctrl, \textbf{2} data.}
%     \label{fig:cc:2}
%   \end{subfigure}
%   \begin{subfigure}{.47\linewidth}
%     \centering
%     \includegraphics[width=\linewidth]{figs/text/yuan/cc3.pdf}
%     \caption{\scriptsize Cross-chiplet flow: \textbf{0} ctrl, \textbf{0} data.}
%     \label{fig:cc:3}
%   \end{subfigure}
%   \caption{Comparison of cross-chiplet traffic in locally-shared LLC (a)---(b) and globally-shared LLC (c)---(d).}
%   \label{fig:cc}
% \end{figure}

\begin{figure}[t]
   \begin{center}
     \includegraphics[width=0.47\textwidth]{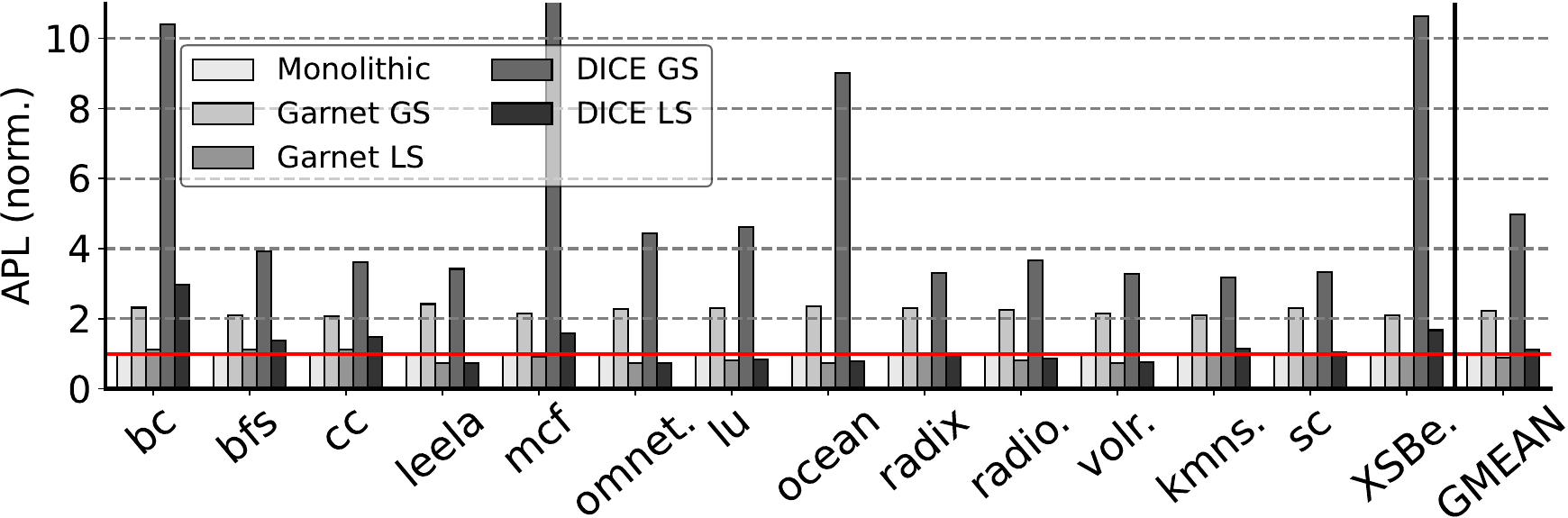}
   \end{center}
   \caption{Normalized APL in global- \textit{vs.} local-shared LLC.}
   \label{fig:APL_IC_CC}
   % \tv{}
\end{figure}

 \begin{figure}[t]
   \begin{center}
     \includegraphics[width=0.47\textwidth]{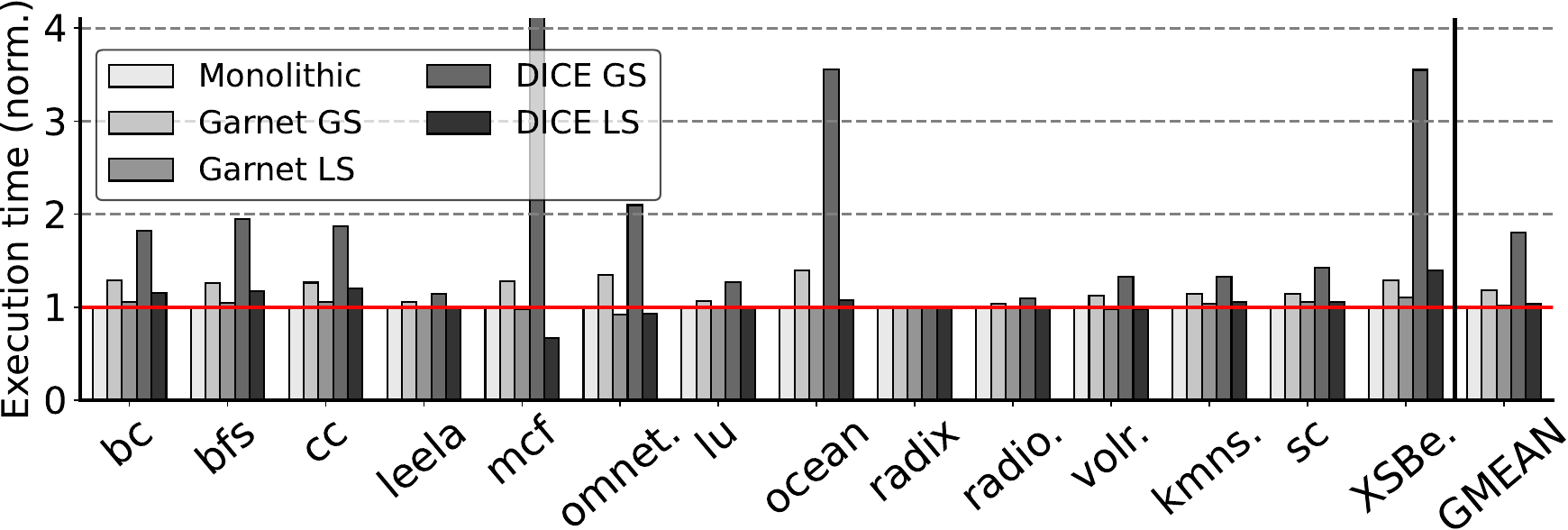}
   \end{center}
   \caption{Normalized exe. time in global- \textit{vs.} local-shared LLC.}
   \label{fig:EXE_IC_CC}
   \tv{}
\end{figure}

% \begin{lstlisting}[language=C++]
%   /* RubySlicc_ComponentMapping */
%   int bits_chiplet; //# of cores in CCD
%   int bits_cpu; // # of cores in CPU 
%   int core_id; // ID number of core
%   bool ic_cc_enable;
%   if (ic_cc_enable){
%       bits_chiplet = bits_cpu;
%   }
%   int chiplet_id = floor(core_id / pow(2,bits_chiplet))
%   machine_number = bitSelect() + (1 << bits_chiplet) * cluster_id
% \end{lstlisting}

\subsubsection{Global vs. Local LLC}
%\textbf{} 
%Finally, to demonstrate the value of \name{} in architectural-level studies we evaluate the trade-offs of sharing the LLC across chiplets. In essence, a chiplet processor is a mini-NUMA system, opening opportunities for trade-offs among coherence scope, interconnect complexity, and scalability.
%
Finally, to demonstrate the value of \name{} in architectural-level studies, we evaluate the trade-offs of sharing the LLC across chiplets. In essence, a chiplet-based processor behaves like a miniature NUMA system, creating opportunities to explore trade-offs among coherence scope, interconnect complexity, and scalability.
For example, AMD EPYC does not maintain cross-CCD coherence~\cite{AMD_EPYC}; each chiplet's LLC is private rather than globally shared---a choice simplifies design/test and preserves modularity. 
%However, the downside is data-sharing inefficiency: as shown in~\autoref{fig:cc:0} and~\autoref{fig:cc:1}, when core C0 in CCD0 incurs a write miss to a cacheline resident in CCD1, the remote copy must be invalidated and the line migrated through the IOD to CCD0; subsequent accesses from CCD1 then need to pull the line back, inducing \textit{ping-pong data-shuffling}, adding latency and extra traffic across chiplet boundaries.
%
On the other hand, architectures such as Intel's Sapphire Rapids~\cite{Intel_sapphire} and AMD's 3D V-Cache~\cite{9563036} employ a globally shared LLC.
%, which eliminates the ping-pong shuffling as shown in~\autoref{fig:cc:2} and~\autoref{fig:cc:3}. A write miss from C0 acquires first write permission and then directly updates the remote LLC in CCD1. In this way, later reads from C1 hits the up-to-date copy \textit{locally}. However, this approach demands a more sophisticated coherence directory to track all cachelines across chiplets, increasing design complexity, verification overhead, and increased chip area.

% Beyond these two extremes, several intermediate coherence optimizations have been proposed. For example, directory-based schemes~\cite{4147673} maintain directory entries that reference remote data, allowing subsequent readers to locate and reuse cached copies without invalidation. SARC coherence~\cite{5582068} predicts write ownership to prefetch read requests efficiently, while the 1-update protocol~\cite{9563036} performs predictive self-downgrades to mitigate remote read stalls.

\red{To facilitate comparative analysis of these 
%coherence 
designs, \name{} provides a configurable %coherence 
scope to switch between globally-shared LLC (GS) and locally-shared LLC (LS), allowing a systematic exploration of chiplet performance and associated design trade-offs.
%allowing a systematic exploration of potential performance benefits and associated design trade-offs .
%is a systematic and accurate way.
%within a controlled simulation environment.
\autoref{fig:APL_IC_CC} and \autoref{fig:EXE_IC_CC} show the effects on packet latency and performance respectively, of globally- or locally sharing the LLC for \emph{multi-programmed workloads} (one copy of each benchmark is run on every chiplet). 
Multi-threaded workloads are beyond the scope of this paper as the coherence effects of locally or globally sharing the LLC warrant a separate study as evidenced in other similar work~\cite{ChengHPCA2007,9563036}. 
As is evident, a GS LLC introduces higher packet latency for both \name{} and \hg{}. The discrepancy between \name{} and \hg{} varies significantly between benchmarks with the resulting \name{} GS performance being less than the monolithic and \hg{} cases. In contrast, both packet latency and performance with an LS LLC are comparable to the monolithic case, and for some benchmarks better because of the smaller, hence faster, on-die chiplet network.
}

\begin{figure}[!t]
   \centering
    \includegraphics[width=0.47\textwidth]{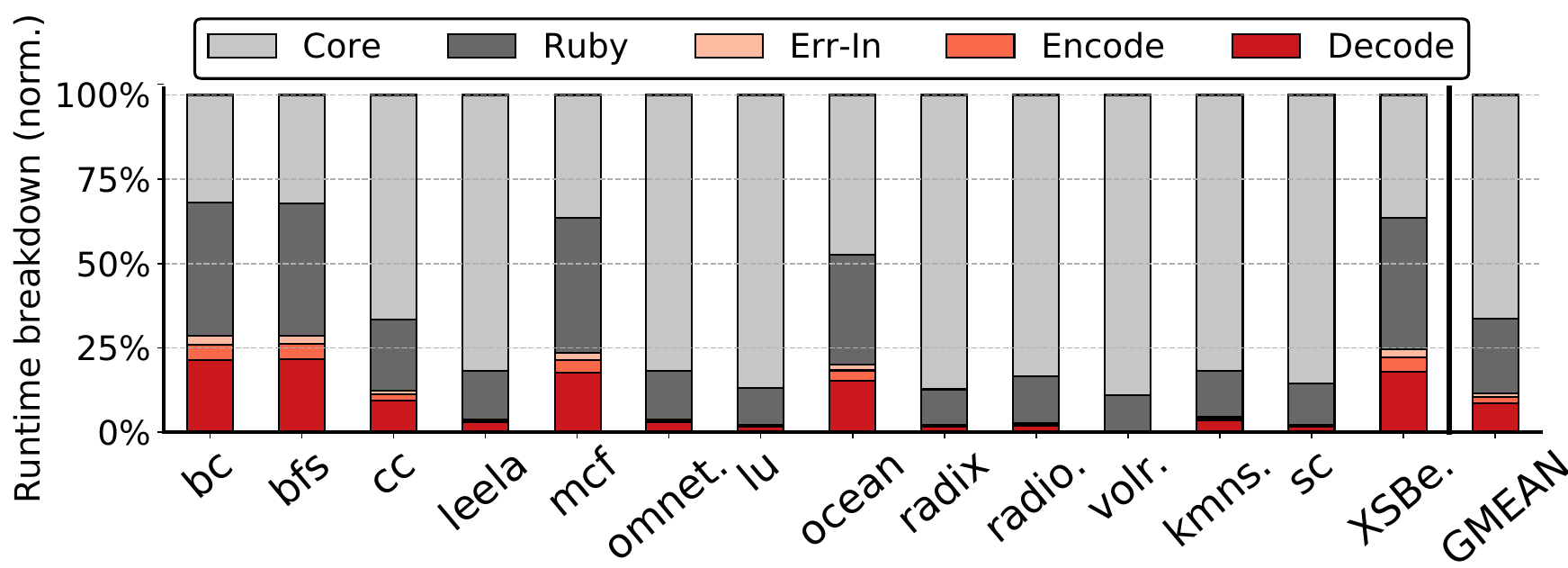}
    \caption{{\color{rebut} 
    % \review{\textbf{[NEW!]}}
    gem5 runtime breakdown, where \textsf{Core+Ruby} corresponds to \textsf{HeteroGarnet}, and the additional overhead introduced by \name{} is highlighted in shades of red.}
    }
    \label{fig:gem5-runtime}
   \tv{}
\end{figure}

{\color{rebut}
\subsection{\name{} gem5 runtime overhead}

\review{B.Q3, E.Q2, F.Q1, F.Q2}

To better understand the runtime overhead of \name{}, we profile gem5 using Linux \textsf{perf\_event} following a methodology presented at the gem5 Workshop at ISCA'25~\cite{gem5-prof}. Profiling runs in a separate process and is non-intrusive, ensuring that it does not perturb gem5's execution time.

\autoref{fig:gem5-runtime} presents the normalized gem5 runtime breakdown at 35\,dB $\mathrm{SNR}_{\mathrm{base}}$, showing the percentage of execution time spent in the \textsf{Out-of-Order Core} and \textsf{Memory (Ruby)}---which together correspond to the runtime of \hg{}---as well as the additional \name{} components, including error injection (\textsf{Err-In}), FEC encoding (\textsf{Encode}), and FEC decoding (\textsf{Decode}).
Across all benchmarks, \name{} introduces limited overhead (\dc{0.3-26.1\%}, averaging \dc{9.2\%}), depending on the application and the amount of inter-chiplet traffic, demonstrating an effective balance between detailed inter-chiplet \phy{}-link modeling and simulation efficiency.

We further observe that the majority of overhead in \name{} is attributed to the layered Min-Sum FEC decoding (\autoref{sec:fec:decode}), which computes iteration-based signal confidence. 
While this is necessary to reflect hardware-level decoding dynamics, \emph{memoization} can be employed, as a future optimization, to cache previously seen symbol patterns and bypass redundant iterative LLR computations, %further 
thus significantly reducing FEC decoding %runtime.
overhead.
%This cost however is necessary to reflect hardware-level decoding dynamics rather than simulator inefficiency. In a future optimization, \emph{memoization} can be employed to cache previously seen symbol patterns and bypass redundant iterative LLR computations, %further 
%reducing FEC decoding %runtime.
%overhead.
}

\begin{figure}[!t]
  \centering
  \begin{subfigure}{0.475\textwidth}
    \centering
    \includegraphics[width=\textwidth]{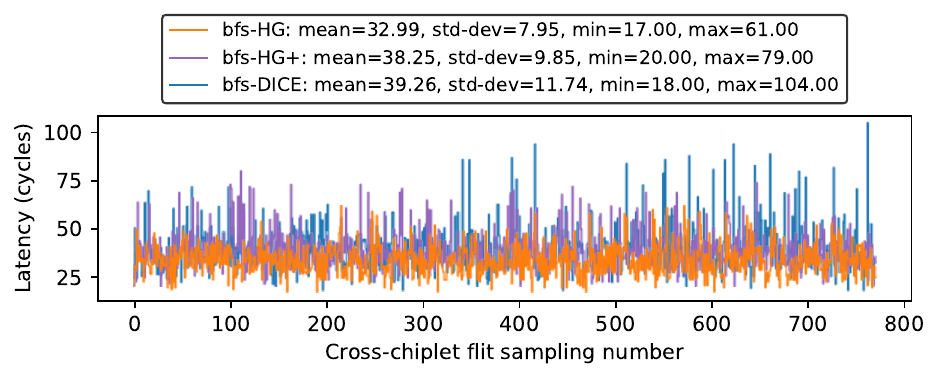}
    \caption{Flit latency sampling (\textsf{bfs})}
    \label{fig:sampling-bfs}
  \end{subfigure} \\
  \begin{subfigure}{0.24\textwidth}
    \centering
    \includegraphics[width=0.88\textwidth]{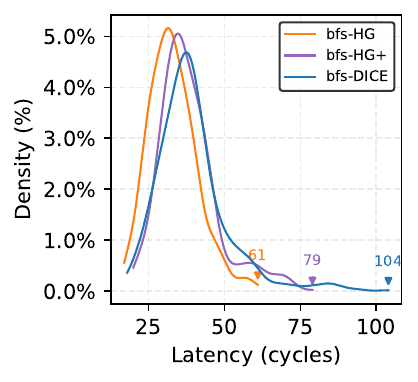}
    \caption{Flit latency histogram (\textsf{bfs})}
    \label{fig:lldist-bfs}
  \end{subfigure}
  \begin{subfigure}{0.24\textwidth}
    \centering
    \includegraphics[width=0.9\textwidth]{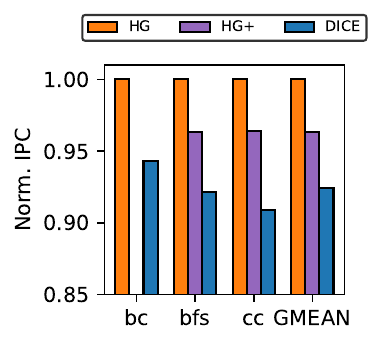}
    \caption{Performance (norm. to \hg{})}
    \label{fig:performance-bfs}
  \end{subfigure}
  \caption{\color{rebut}
  % \review{\textbf{[NEW!]}}
  \hg{}, \hg{}+, and \name{}: Latency variability and resulting IPC for 3 high-MPKI workloads. \textit{Note:} In \textsf{bc}, \hg{}+ induces long backlogs that ultimately lead to simulation failure.
  }
  \label{fig:variability-bfs}
  % \tv{}
\end{figure}

{\color{rebut}
%\section{Discussion: Architectural Implications}
%\section{Discussion: Architectural Relevance}
\subsection{Architectural relevance}
\label{sec:relevance}
\review{A.Q1, C.Q1, E.Q1, F.Q3}

%\textbf{For ISCA's audience.} 
Our motivation for \name{} %closely 
parallels the justification for detailed DRAM-system simulators where the key point is not that average memory latency and bandwidth are unknown; it is that a static ``latency + bandwidth'' abstraction is too coarse to predict performance. This is because the performance-relevant quantity is the dynamic interaction between 1) the workload's access/communication pattern and 2) the microarchitectural mechanisms that schedule, queue, overlap, and serialize those events~\cite{ghose2019demystifying}. This interaction produces high run-time variability in effective memory service times, and it is precisely that \textit{variability}, rather than a single \textit{mean}, that drives stalls, contention, and an increase in critical-paths~\cite{ghose2019demystifying}. 

\textbf{\name{}'s relevance to OoO microarchitecture.} Chiplet-based systems create a similar situation for core-to-core or core-to-IOD communication where data transfers %and coherence traffic 
experience latency and variability that can be comparable to DRAM. As a result, treating inter-chiplet communication as a fixed constant has the same failure mode as constant DRAM models---it erases the variability that determines when (or how) latency affects performance.

 More specifically, modern OoO cores can tolerate a bounded amount of load/store latency through memory-level parallelism, speculation, and overlap. When a latency stays below this ``absorbable'' region, it may be largely hidden; when it exceeds it, it becomes increasingly visible as pipeline stalls, ROB and LSQ pressure, and exposed dependence stalls. This is best explained by the first order models of Karkhanis and Smith~\cite{karkhanis2004first}, and Eyerman \textit{et al.}~\cite{eyerman2009mechanistic}.
A constant-latency model therefore forces an inaccurate outcome:
If the chosen constant is ``low enough,'' the simulation 
%will predict that inter-chiplet communication has essentially no 
ignores the performance impact (because OoO hides it). If the chosen constant is ``high,'' the simulation yields a systematic, \textit{uniform} slowdown. 
In both cases, it fails to capture latency variability and worst-case scenarios, resulting in simulation inaccuracies.
%Thus, there is no single defensible constant for inter-chiplet latency. 
This is corroborated by what we observe with \name{}.
\autoref{fig:variability-bfs} shows 1) sampled flit latencies for cross-chiplet loads in \textsf{\dc{bfs}} collected every 50,000 cross-chiplet flits (35\,dB $\mathrm{SNR}_{\mathrm{base}}$), 2) the corresponding latency histogram, and 3) the impact of \name{} on performance for \textsf{\dc{bfs}} and two other high-MPKI benchmarks, \textsf{\dc{bc}} and \textsf{\dc{cc}}.
% Observe that the resulting difference in performance between \name{} and \hg{} for \textsf{\dc{bfs}} cannot be justified by the meager difference in their latency means ($\approx$4 cycles, \autoref{fig:sampling-bfs}), but rather is a result of the large \emph{difference of the tail latencies} (max: \dc{102} cycles for \hg{}, \dc{163} for \name{}, \autoref{fig:sampling-bfs}).

We can observe that the performance gap between \name{} and \hg{} cannot be justified by the meager difference in their average-latency shown in~\autoref{fig:sampling-bfs} ($\approx$6 cycles: \dc{32.99} for \name{} \vs{} \dc{39.26} for \hg{}), but rather is a result of the large \emph{difference of the tail latencies} (\dc{61} cycles for \hg{} \vs{} \dc{104} cycles for \name{}).
Further, statically increasing \hg{}'s cross-chiplet link-cycle (denoted \textsf{bfs}-\hg{}+, reaching an average-latency of \dc{38.25} cycles) to match \name{}'s average-latency affects application performance only to a limited extent (\autoref{fig:performance-bfs}), as it aligns the average but still leaves a substantial gap in the long-tail (\autoref{fig:lldist-bfs}).
Further, such adjustments also vary across applications, rendering the tuning process inherently \textit{ad hoc}. 
In some cases (\eg{}, \textsf{bc}), \hg{}+ induces severe request backlogs and can even cause simulation failure.
Overall, these results show that, for OoO microarchitecture, \hg{}/\hg{}+ with constant link-latency are insufficient to reproduce \name{}.
}

%The exposed latency depends on dynamic link utilization, coherence traffic, queue occupancy, and endpoint contention; the same workload can experience periods where messages are absorbed and periods where they affect the critical path.

%This is exactly where \name{} changes what can be studied: by modeling latency variability over time, it can capture when inter-chiplet transactions are (a) overlapped and benign versus (b) serialized on dependence chains and performance-critical.

%Analogously, 
%Similarly, 

\begin{figure}[!t]
  % \begin{subfigure}{0.24\textwidth}
  \centering
  % \includegraphics[width=1.1\textwidth]{figs/isca_rebuttal/hg-vs-DICE-multi/res/loadToUse_stats_compare.pdf}
  % \caption{Load statistics (norm. to \hg{})}
  \includegraphics[width=0.475\textwidth]{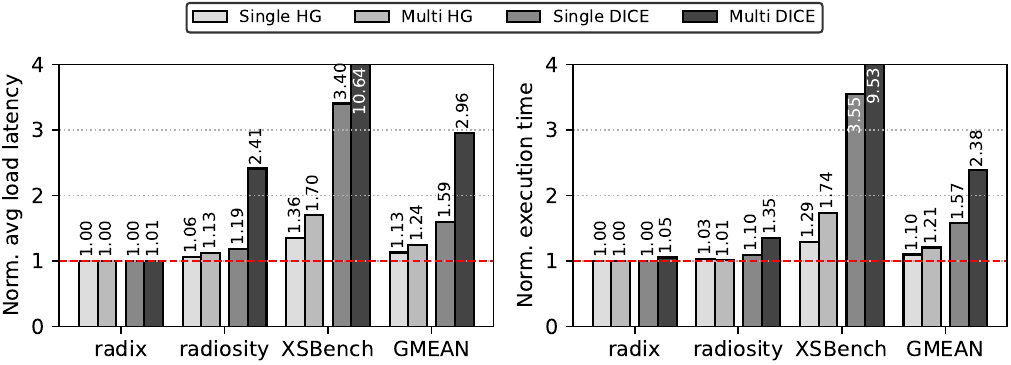}
  \caption{\color{rebut}
  %Average load latency (left) and application execution time (right), all normalized to \textsf{Monolithic}, for three multi-threaded workloads.
  Single- vs. multi-threaded: Load latency (left) and application execution time (right), normalized to \textsf{Monolithic}.
  }
  \label{fig:multithreading}
  \tv{}
  % \label{fig:multithreading:ltu}
  % \end{subfigure}
  % \begin{subfigure}{0.24\textwidth}
  %   \centering
  %   \includegraphics[width=0.9\textwidth]{figs/isca_rebuttal/hg-vs-DICE-multi/res/simTick.pdf}
  %   \caption{Performance (norm. to \hg{})}
  %   \label{fig:multithreading:exec}
  % \end{subfigure}
  % \caption{\color{rebut}
  % \review{\textbf{[NEW!]}}
  % Load latency statistics and application execution time under \name{}, normalized to \hg{} (dashed line), for three multi-threaded workloads.
  % }
\end{figure}

{\color{rebut}
%\textbf{\name{}'s impact on multi-threading.}
\review{B.Other}
\textbf{\name{}'s relevance to coherence and synchronization.}
In the same vein, for coherence traffic that gates lock transfers, barriers, hand-offs, and producer–consumer relationships,
%. In these cases, 
tail latency becomes even more critical because forward progress is determined by the slowest coherence transaction on the synchronization path. %This is a well-known phenomenon in synchronization: 
Small changes in the latency of lock release/acquire can produce disproportionate end-to-end effects, as hand-off latency repeats many times (serialized) and stretches the serial fraction of execution, which disproportionally affects performance (Amdahl's Law). 
Prior work has indeed shown the sensitivity of lock performance to packet latency and backoff behavior~\cite{ros2015callback}; that same sensitivity becomes more acute when coherence messages traverse inter-chiplet links whose delay is both higher and more variable than on-die.

To highlight \name{}'s impact on multi-threaded programs, we run three benchmarks: \textsf{radix}, a lightly synchronized workload~\cite{splash4}; \textsf{radiosity}, a synchronization-intensive workload~\cite{splash4}; and \textsf{XSBench}, a heavily synchronized workload with a large memory footprint~\cite{xsbench}. We use 4 CCDs $\times$ 8 cores (32 cores total) with a globally shared LLC, which, as shown in \autoref{fig:APL_IC_CC}, can significantly affect average packet latency. %(a proxy for average load latency) in some benchmarks. 
%In \autoref{fig:multithreading}, right graph, we juxtapose the increase in execution time of the single-threaded and multi-threaded versions of these benchmarks, for \hg{} and \name{} (all normalized to \textsf{Monolithic}). \autoref{fig:multithreading}, left graph, shows the corresponding increases in load latency. 
In \autoref{fig:multithreading}, the right graph juxtaposes the increase in execution time for the single-threaded and multi-threaded versions of these benchmarks, for \hg{} and \name{} (all normalized to \textsf{Monolithic}). The left graph shows the corresponding increases in load latency.
As the figure shows, both execution time and average load latency increase sharply in the multi-threaded case, with the magnitude of the increase depending on 1) the benchmark's synchronization and memory intensity, increasing from \textsf{radix} to \textsf{XSBench}, and 2) the fidelity of the interconnect model, increasing from \hg{} to \name{}.
For the multi-threaded \textsf{XSBench} under \name{}, execution time rises to \dc{9.53$\times$} that of \textsf{Monolithic}, versus \dc{3.55$\times$} for the single-threaded case, while under \hg{}, the corresponding slowdowns are only \dc{1.74$\times$} (multi-) and \dc{1.29$\times$} (single-threaded).

\textbf{Takeaway.} Accurate chiplet-communication modeling is essential for studies of OoO cores, cache coherence, and synchronization. Because its effects are strongly workload dependent, a ``constant'' latency proxy is unreliable. The largest impact appears in workloads with long dependent-load chains and little latency tolerance, and in coherence/synchronization-heavy kernels, where latency variability can trigger an inordinate number of stalls on relatively few critical long-latency messages.

}

%%%%%%%%%%%%%%%%%%%%%%%%%%%%%%%%%%%%%%%%%%%%%%%%%

\section{Conclusion}

% Chiplet-based architectures are rapidly becoming the foundation of scalable, high-performance processors. As inter-chiplet communication continues to scale with denser wiring, higher signaling rates, and more aggressive integration, channels are pushed closer to their limits, making transmissions less reliable and more susceptible to errors. However, this design space is often under-evaluated, as state-of-the-art simulators typically oversimplify inter-chiplet links, modeling \phy{} as an offline, fixed constant.

% In this work, we show that \phy{}-level behavior introduces meaningful variability in both packet timing and overall system performance. To address this gap, we propose \name{}, which integrates calibrated, runtime-accurate \phy{} modeling into \textsf{gem5}, capturing FEC encoding/decoding, modulation, lossy-channel effects, adaptive retransmissions, and fine-grained inter-chiplet flow control. Through extensive evaluation, we demonstrate that realistic \phy{} dynamics reshape packet-latency distributions and shift system-level metrics such as IPC by up to \dc{27.6\%}, exposing trends that remain invisible under fixed-latency models and enabling more faithful design exploration for next-generation chiplet systems.

In this work, we show that \phy{}-level behavior introduces meaningful variability in both packet timing and overall system performance. To address this gap, we propose \name{}, which integrates calibrated, runtime-accurate \phy{} modeling into \textsf{gem5}. Through extensive evaluation, we demonstrate that realistic \phy{} dynamics reshape packet-delay distributions and shift system-level metrics such as IPC by up to \dc{27.6\%}. Our analysis further reveals that packet variability and long-tail delays---rather than average latency---play a critical role in system performance. These effects are difficult to capture with prior simulators that rely on constant link-latency models, highlighting the importance of \phy{}-accurate modeling for faithful chiplet-system design exploration.

\section*{Acknowledgments}
This work is supported by the Swedish Research Council (VR) starting grant number 2025-04436, by the VR grant 2022-04959, and by the Swedish Foundation for Strategic Research (SSF) grant FUS21-0067. We acknowledge the use of computing resources provided by the National Academic Infrastructure for Supercomputing in Sweden (NAISS) under project numbers NAISS 2024/6-189 and NAISS 2025/2-430.

\balance
\bibliographystyle{ieeetr}
\bibliography{papers}

\end{document}